\documentclass{article}

\usepackage{etoolbox}
\usepackage{graphicx} 
\usepackage{grffile}
\usepackage{subfigure, amsmath, amssymb}
\usepackage{amsthm} 
\usepackage{algorithm}
\usepackage{algpseudocode}
\usepackage{balance}
\usepackage{color}
\usepackage{url}
\usepackage{mathtools}
\usepackage{paralist}

\usepackage{cite}
\usepackage{amsfonts}
\usepackage{textcomp}
\usepackage{xcolor}

\providecommand{\keywords}[1]
{
  \small	
  \textbf{\textit{Keywords---}} #1
}

\begin{document}

\title{Automated Trajectory Synthesis for UAV Swarms Based on Resilient Data Collection Objectives}

\author{A H M Jakaria, Mohammad Ashiqur Rahman, \and Matthew Anderson, and Steven Drager}

\maketitle
\begin{abstract}
The use of Unmanned Aerial Vehicles (UAVs) for collecting data from remotely located sensor systems is emerging. The data can be time-sensitive and require to be transmitted to a data processing center. However, planning the trajectory of a collaborative UAV swarm depends on multi-fold constraints, such as data collection requirements, UAV maneuvering capacity, and budget limitation. Since a UAV may fail or be compromised, it is important to provide necessary resilience to such contingencies, thus ensuring data security. It is important to provide the UAVs with efficient spatio-temporal trajectories so that they can efficiently cover necessary data sources. In this work, we present Synth4UAV, a formal approach for automated synthesis of efficient trajectories for a UAV swarm by logically modeling the aerial space and data point topology, UAV moves, and associated constraints in terms of the turning and climbing angle, fuel usage, data collection point coverage, data freshness, and resiliency properties. We use efficient, logical formulas to encode and solve the complex model. The solution to the model provides the routing and maneuvering plan for each UAV, including the time to visit the points on the paths and corresponding fuel usage such that the necessary data points are visited while satisfying the resiliency requirements. We evaluate the proposed trajectory synthesizer, and the results show that the relationship among different parameters follow the requirements while the tool scales well with the problem size.
\end{abstract}


\section{Introduction}
\label{Sec:Introduction}

Unmanned Aerial Vehicles (UAVs) are increasingly being used for data collection. They are successfully used in scenarios where it is highly dangerous, as well as monotonous for human observers, or in some cases, excessively costly~\cite{jakaria2018formal}. For example, UAVs may be used in surveillance of sensitive points in industrial networks, such as the smart grid bus system~\cite{deng2014unmanned}. They can also be used in UAV-assisted wireless sensor networks to collect data directly from sensor nodes~\cite{abdulla2014optimal}. Sensor nodes can be planted at prespecified remote locations on the ground. Some of the UAV waypoints can be located in close proximity to the data sensors. 
In many cases, UAVs start their flights from one or more base stations or data centers and collect data from different waypoints along the trajectories before reaching endpoints. 

The trajectory planning in scenarios that consider multi-fold constraints of the UAVs, such as maneuvering restriction, budget limitation, and data collection requirements, is generally an NP-hard problem~\cite{yeung2016routing}. The trajectories of UAVs may contain predefined waypoints in the three-dimensional Euclidean space; however, the UAVs have certain limitations in maneuvering and fuel capacity while going from one point to another. The operators of such data collection systems generally employ a swarm of UAVs to collect data from the data sources collaboratively but may have budget limitations. The data collection is time-sensitive, as data freshness is important in practical scenarios. For example, the operator may want a data point to be resiliently covered by multiple UAVs, so that even if some of the UAVs fail to operate normally due to malfunction or cyberattacks, some others can still perform the job successfully. Moreover, they need to make sure that the difference in the time of the data collection is below certain thresholds. 
This calls the need for a comprehensive UAV trajectory designer for a UAV swarm, although the problem is combinatorially hard.

In this paper, we aim to address this challenge by developing Synth4UAV, a formal model-based trajectory planner for a UAV swarm. We use Satisfiability Modulo Theories (SMT)~\cite{de2011satisfiability} to implement, encode, and solve the corresponding formal model. The solution synthesizes a trajectory for the UAV swarm, if there is one, satisfying all the given constraints, including data freshness and budget constraints. The evaluation results show that our tool scales well to solve problems with large spatial systems. 
Our major contributions in this research consist of the following:
\begin{compactenum}
  \item We formally model the UAV trajectory planning as a constraint satisfaction problem. We model the UAV moves and associated constraints, such as UAV maneuver limitations, fuel usage, data freshness, budget limitation, and resiliency requirements for data collection.
  \item We encode the formal model using SMT as a constraint satisfaction problem and solve it using an efficient SMT solver to find the trajectory of the UAV swarm. We demonstrate the developed tool on a case study.
  \item We analyze various characteristics of the trajectory synthesis problem. We evaluate the scalability of the tool. We build a graphical simulator for visual analysis.
\end{compactenum}

The rest of this paper is organized as follows: Section~\ref{Sec:Background} presents the research problems, challenges, and related works. We discuss the framework of our developed solution in Section~\ref{Sec:Framework}. Section~\ref{Sec:Model} describes the formal model of the constraints and requirements. The implementation details and a case study are presented in Section~\ref{Sec:Imp}. The evaluation results of the implemented tool is presented in Section~\ref{Sec:Eval}.  Finally, we conclude the paper in Section~\ref{Sec:Conclusion}.
\section{Background}
\label{Sec:Background}

In this section, we discuss the research problem, the challenges associated with it, and the related works.

\subsection{Research Problem}
A UAV swarm is a network of UAVs that can communicate with each other (mostly following the ad-hoc wireless networking) and usually perform a common task, autonomously or with the instructions from operators. It is important to maintain appropriate trajectories for all the UAVs in a network in order to avoid mid-air-collision or restricted areas ({\it e.g.}, adverse territory or restricted private locations). It is also required to maintain successful communication between the neighboring UAVs and cover the surveillance points within a time constraint. In this research, we propose a formal framework that plans for the trajectories of the UAVs in a swarm, so that all or at least a fraction of the points near the data sources are covered within a given period. The coverage requirement may ensure that at least one of the available UAVs should cover each data point once or at each operator-specified time interval within the period. The resiliency specification considered in this research states that even if a certain number ($r$ or $k$) of UAVs fail to operate due to technical failures or cyberattacks, the surveillance goal is still fulfilled. In other words, the trajectory plan for the remaining UAVs can still securely surveil the required percentage of the data points.

\begin{figure}
	\centering
	\includegraphics[keepaspectratio,scale=0.21,bb=0 0 1809 708]{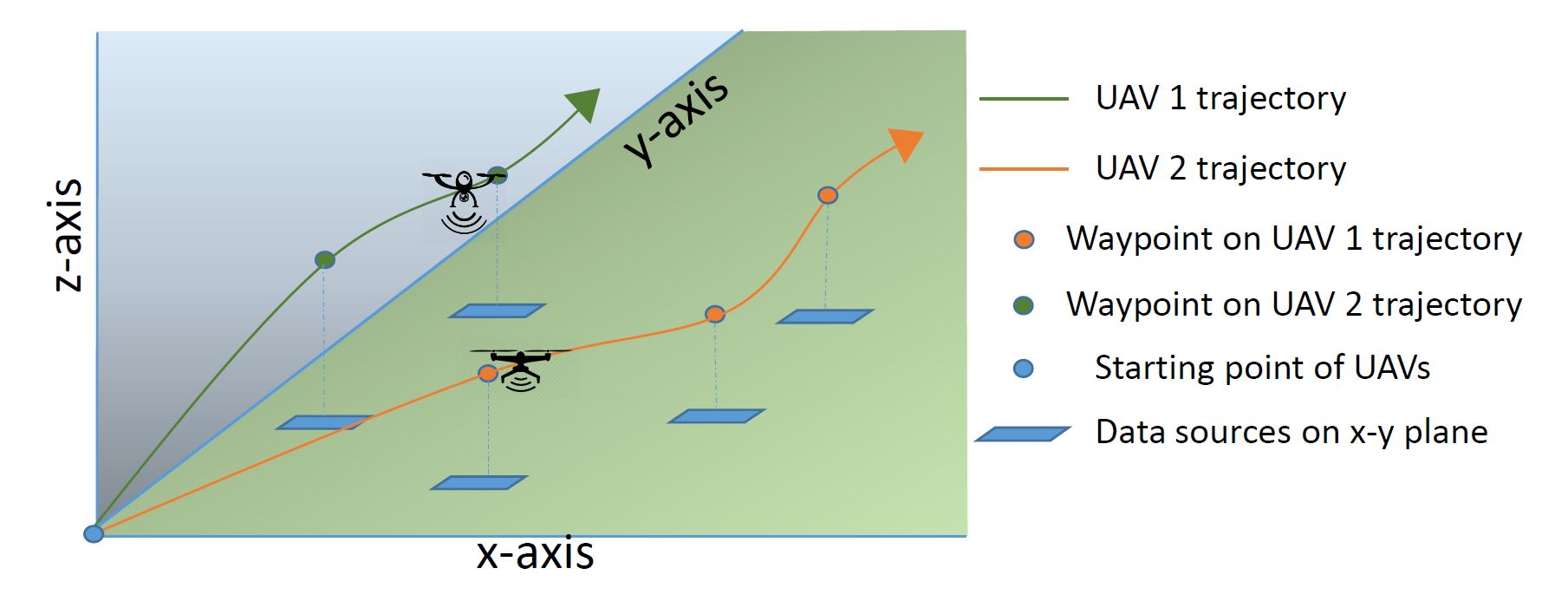}
	\vspace{-12pt}
	\caption{Sample trajectories of two UAVs.}	
	\label{Fig_Traj}
	\vspace{-6pt}
\end{figure}

Fig.~\ref{Fig_Traj} presents two sample trajectories for two different UAVs. They collect the available data in a combined way. The figure shows the starting points for both the UAVs, as well as the positions or waypoints from where they can collect the data. In our research problem, we consider a set of points in the three-dimensional space where the UAVs can travel, as well as subsets of the points where data is available and where traveling is forbidden.

\subsection{Research Challenges} 
A swarm of UAVs is deployed to collect data over an area from a set of predefined sensors planted on the ground. The sensors are located on the $XY$ plane of the three-dimensional Euclidean space with $X$, $Y$, and $Z$ axes. There are multiple points with $xyz$ coordinates, of which some are directly above the data sensors or near to them from where the UAVs can collect the transmitted data. Higher $z$ value denotes points with higher altitude. A subset of the points are restricted, which must be avoided during the flight. Designing trajectories of all the UAVs so that the budget and time restrictions are met, they do not collide with one another, avoid restricted areas, and still perform the desired data collection surveillance is a multi-objective problem (NP-hard). We use satisfiability modulo theories (SMT) to solve this problem and generate the trajectories in the 3D space that satisfies all the constraints and requirements.

\subsection{Related Works}
There are several works in the literature that provides solutions to the problems of path projection and path planning for UAVs. For example, Tisdale {\it et al.} proposed a mechanism to perform path planning while minimizing the uncertainty in sensing missions by considering a blend of online sensor models and estimation objectives~\cite{tisdale2009autonomous}. 
Yeung {\it et al.} proposed two heuristic-based techniques to find an optimal route of aerial vehicles so that the original target and waypoints are not affected~\cite{yeung2016routing, yeung2015trajectory}. They also consider the flight maneuvering constraints of the UAVs and keep the deviation from the original route to a minimum while maximizing the number of sensor tasks from the ground. They also consider the time deadlines to cover the taskpoints and maintenance of QoS. Mekala {\it et al.} proposed a similar method to generate the flight trajectory, which is a collection of predefined waypoints~\cite{mekala2016aerial}. They also propose an algorithm to optimize the flight path. 
Liu {\it et al.} studied the age-optimal trajectory planning in UAV-enabled wireless sensor networks~\cite{liu2018age}. They ensure that the age of the data collected from different sensor nodes is bounded by a threshold. They proposed a dynamic programming method and a genetic algorithm-based method.

Beard {\it et al.} designed semi-autonomous UAVs and proposed their graph-based approach to path planning and trajectory generation~\cite{beard2005autonomous}. Ceccarelli {\it et al.} developed a path planning model for micro UAVs to obtain video footage of a set of known targets~\cite{ceccarelli2007micro}. In the presence of constant winds, they find the best selection of waypoints for achieving the reconnaissance goals. 
A pheromone model, where the movement of a group of UAVs is guided by pheromone, was proposed by Kuiper {\it et al.}~\cite{kuiper2006mobility}. This model performs well in reconnaissance.

A two-step algorithm was proposed by Bortoff to obtain a path for UAVs traveling through unauthorized sites~\cite{bortoff2000path}. The algorithm performed a trade-off between the shortest path for traveling and avoiding radars.
A trajectory design with the aid of an iterative algorithm was proposed by Wu {\it et al.} where aerial base stations for a wireless communication system on the ground were mounted on the UAVs~\cite{wu2018joint}. 
Bagherian proposed a genetic and particle swarm algorithm to generate the optimal three-dimensional path planning in an adverse environment~\cite{bagherian20153d}. They considered the terrain information and the kinematic constraints of the UAVs.

Tsourdos {\it et al.} presented a Kripke model for UAV path planning to cope with uncertainties and decision making involved in the process of collaboratively reaching the goal while covering a certain area of interest, avoiding obstacles~\cite{tsourdos2005formal}. They modeled the collision avoidance with the help of conditions such as minimum separation and non-intersection of paths at equal length. Their model can find the shortest path for each UAV. 
Ruz {\it et al.} proposed a Mixed Integer Linear (MILP)-based solution which modifies the A* algorithm to optimize paths in dynamic environments where pop-ups appear according to a known future probability and need to be avoided~\cite{ruz2007decision}. Their ILP model can choose the most suitable trajectory among different alternatives provided fuel and time constraints. 
The authors also presented a similar approach to trajectory optimization for UAVs in the presence of obstacles, waypoints, and risk zones in another research~\cite{ruz2009uav}. 
Jakaria {\it et al.} presented a formal model for verification of a resilient communication in a UAV swarm while maintaining safe flight trajectories for all the UAVs~\cite{jakaria2018formal, jakaria2018safety}.
Jameson proposed a fuel consumption algorithm for UAVs that addresses the additional fuel requirements issue in the Aviation Weather Routing Tool (AWRT) developed by the Army Research Laboratory~\cite{jameson2009fuel}. They developed a prototype algorithm that computes the ground speeds along the different mission segments, fuel consumed, and fuel remaining while rerouting during adverse weather conditions.

%
%


Chen {\it et al.} optimized the trajectory of quad-rotor UAVs in collaborative environments~\cite{chen2016trajectory}. They proposed a hierarchic optimization technique to solve the trajectory optimization problem of quad-rotor UAVs.
Sun {\it et al.} created a trajectory-tracking controller where the autopilot is involved in the feedback control loop~\cite{suntrajectory}. They proposed a generalized trajectory design model using Lyapunov based backstepping.
Fadlullah {\it et al.} proposed a simple trajectory control algorithm for UAVs that manages the throughput of UAV-aided networks, where UAVs work as network nodes~\cite{fadlullah2016dynamic}. 
Kvarnstrom {\it et al.} presented a framework for UAV mission planning that combines ideas from forward-chaining planning and partial order planning~\cite{kvarnstrom2010automated}. They met the requirements of centralization, abstraction, and distribution of task assignments. 
Bethke {\it et al.} proposed a health management system for UAV teams that perform persistent surveillance~\cite{bethke2008group}. Their methodology for planning missions can anticipate the negative effects of various types of anomalies on future mission states, as well as choosing actions to mitigate the effects. 

However, none of these works present the path planning for resilient coverage of data points. We consider the resiliency of the planned trajectories, which ensures a certain coverage of the data sources with a certain frequency of timely visits of multiple UAVs. We solve an NP-hard problem~\cite{sahingoz2014generation}, which involves trajectory synthesis while satisfying multi-objective constraints such as budget and maneuvering limitations, and operator requirements, such as resilient coverage.
\section{Synth4UAV Framework}
\label{Sec:Framework}

We develop Synth4UAV, a formal model-based tool that creates a trajectory plan for each UAV of a UAV swarm, where a trajectory is a sequence of waypoints through which the UAV will fly. The trajectory depends on current positions of the UAVs, their velocity, climb angle, and turn angle, etc. Different UAVs may have different speeds, which are assumed to be constant throughout their flight time while performing the surveillance. The UAVs can start from different base stations. For a time period (represented as a series of time steps), the trajectory is modeled. Within a unit time step, the UAVs will not alter their flight parameters. The output trajectory is a set of coordinates in the 3D space, along with the climb angle and turn angle in each step. We model the fuel consumption where a UAV starts traveling with a certain amount of fuel and returns to a base station for refueling if needed. 

\begin{figure}
	\centering
	\includegraphics[keepaspectratio,scale=0.25,bb=0 0 1760 805]{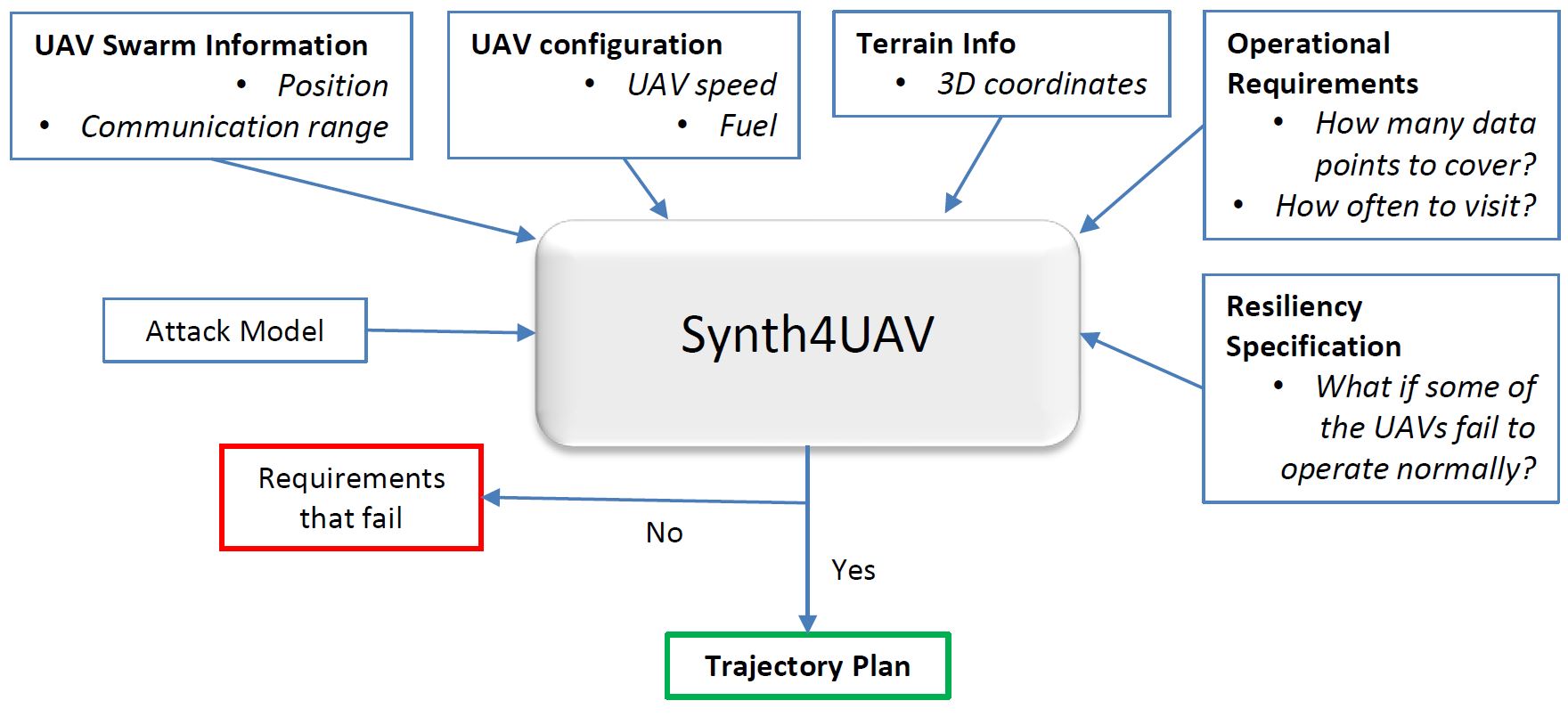}
	\vspace{-7pt}
	\caption{A formal framework for trajectory planning.}
	\label{Fig_Frame}
\end{figure}

Fig.~\ref{Fig_Frame} presents the framework of Synth4UAV. The framework takes necessary inputs, such as the UAV swarm and individual UAV properties, terrain information, and data coverage requirements. The swarm information includes the number of UAVs, initial positions of the UAVs, as well as their speeds and directions. Surveillance requirements specify the points to visit and the characteristics/constraints associated with these visits. The formal model is eventually solved using SMT logic/solver, to identify the trajectory plan ({\it e.g.}, the flight angles), satisfying the requirements and constraints of the surveillance. Our trajectory planning considers necessary invariants. For example, the climb and turn angles must be within the (practical) capacity of a UAV.

\section{Model}
\label{Sec:Model}

This section provides the formal modeling of the requirements and the constraints of the UAV trajectory planning. Table~\ref{Tab_Notation} lists several variables used in the model. It is notable that no multiplication of two parameters is performed in the paper without the multiplication sign.

\subsection{Preliminary of Model}
To model the trajectory planning for the UAVs, we consider a set of predefined points (waypoints) in the three-dimensional space. Each point is considered to have Cartesian ($xyz$) coordinates in the Euclidean space. The UAVs start and end their flights at certain points, which are generally base stations or fueling stations. Between the starting and ending points, the UAVs may visit several other points. All these points constitute the trajectories for all the available UAVs. Our basic intention is to automatically generate these trajectories such that certain constraints and requirement goals are met. Here we formally model the route constraints, as well as the requirements of the users of the implemented tool. We make sure that the points containing the data sensors are covered with certain resiliency specifications. In other words, even if some of the UAVs fail to operate normally to visit certain data points, the points are still covered by others so that the data is collected efficiently.

\begin{table}[t]
\caption{Notation Table} \label{Tab_Notation}
\centering
\small
\begin{tabular}{|p{0.7in}|p{3in}|}
\hline\textbf{Notation} & \textbf{Definition} \\
\hline
$\mathit{M}^{\hat{p}, p}_u$ & Boolean representing movement of UAV $u$ from point $\hat{p}$ to $p$ \\
$\mathit{V}^{p}_u$ & Boolean representing that UAV $u$ has visited point $p$ \\
$\mathit{L}^{\hat{p}}_p$ &  Boolean representing the possibility of traveling from point $\hat{p}$ to $p$ \\
$\mathit{C}^{p}_{u}$ & Real expression denoting the cost of fuel for UAV $u$ for reaching point $p$ \\
$\mathit{T}^{p}_{u}$ & Real expression denoting the time taken by UAV $u$ to reach point $p$ since its starting point \\
$\mathit{H}^{p}_{u}$ & Integer expression denoting the time a UAV $u$ hovers over point $p$ \\
\hline
\end{tabular}
\normalsize
\end{table}

\subsection{Modeling UAV Movement}
\label{subsec:movement}
If $\mathit{M}^{\hat{p}, p}_u$ denotes whether UAV $u$ moves from $\hat{p}$ to $p$, UAV $u$ has already visited point $\hat{p}$. This is denoted by $\mathit{V}^{\hat{p}}_u$. Also it must be possible for the UAVs to travel from point $\hat{p}$ to $p$. In other words, there should be a link between point $\hat{p}$ and point $p$, which is denoted by $\mathit{L}^{\hat{p}}_p$.
\vspace{-3pt}
\begin{equation}
\begin{split}
& \mathit{M}^{\hat{p}, p}_u~\rightarrow~\mathit{V}^{\hat{p}}_u \wedge \mathit{L}^{\hat{p}}_p
\end{split}
\end{equation}

We define the Boolean variable $\mathit{D}^{\hat{p}}_p$ as true only if the distance between point $\hat{p}$ and $p$ ($\mathit{dist}^{\hat{p}}_p$) is within a certain threshold, $d_{u,th}$. The threshold is specified by the properties of UAV $u$.
\vspace{-3pt}
\begin{equation}
\begin{split}
& \mathit{D}^{\hat{p}}_p~\leftrightarrow~\mathit{dist}^{\hat{p}}_p \le d_{u,th}
\end{split}
\end{equation}

\begin{figure}
	\centering
	\includegraphics[keepaspectratio,scale=0.25,bb=0 0 1803 700]{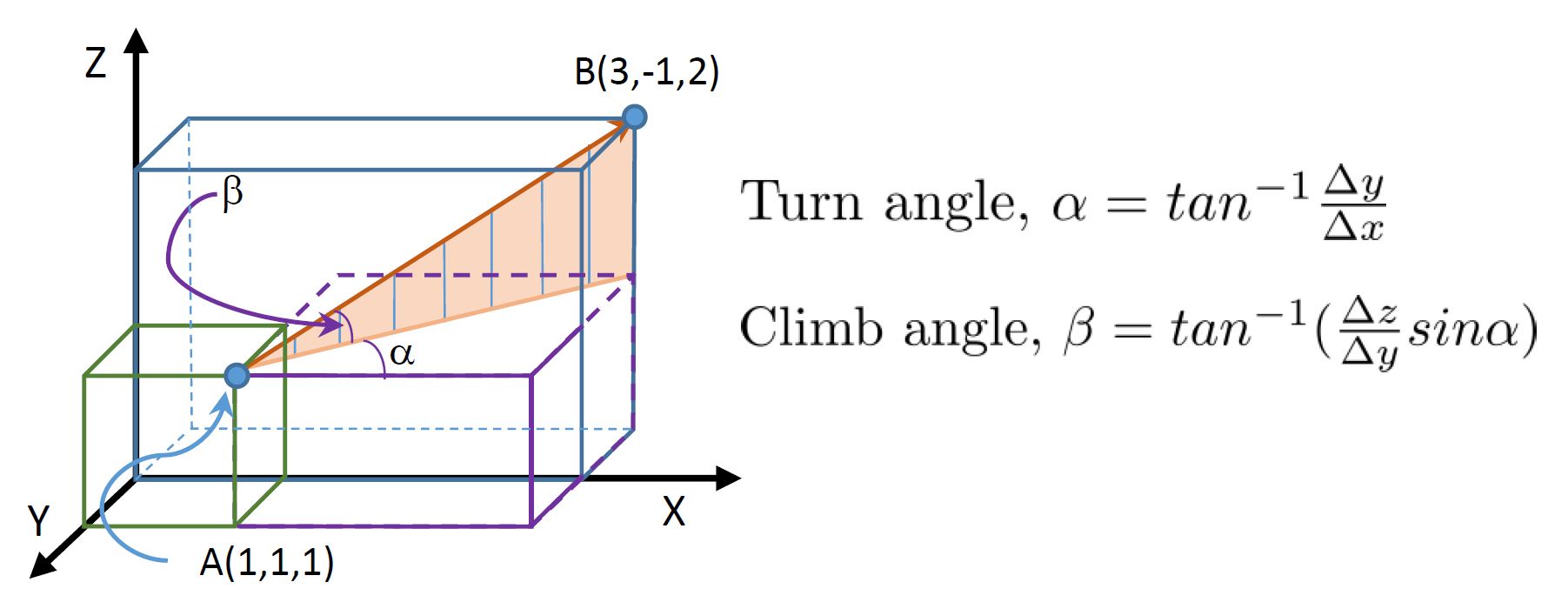}
	\vspace{-9pt}
	\caption{Turn and climb angles while moving from one point to another.}
	\label{Fig_Angle}
\end{figure}

The UAVs have certain maneuvering constraints. We consider the restrictions on the turn angle and the climb angle in three-dimensional space~\cite{bagherian20153d}. The turn angle is the angle that a UAV has to turn with respect to the existing direction on the horizontal plane. We first calculate the angle for a UAV for going from $\hat{p}$ to $p$ and the existing angle which the UAV is following to reach $\hat{p}$ from $p'$. Then we calculate the deviation which the UAV has to turn. Similarly, we calculate the climb angle on the vertical plane. 
Fig.~\ref{Fig_Angle} shows an example scenario of moving from one point to another in the three-dimensional space. $\alpha$ is the turn angle for moving from point $A$ to $B$, provided the direction of a UAV at $A$ is parallel to the $X$ axis. $\beta$ is the climb angle. Fig.~\ref{Fig_Angle} also provides the formulas for calculating the angles based on the horizontal and vertical movements.

The deviation of turn angle to reach $p$ from $\hat{p}$, given a UAV reached $\hat{p}$ from $p'$ is represented by $\delta^{p',\hat{p},p}_{\alpha}$, while the deviation of climb angle to reach $p$ from $\hat{p}$, given it reached $\hat{p}$ from $p'$, is denoted by $\delta^{p',\hat{p},p}_{\beta}$. The deviations must be within certain thresholds ($\theta^{t}_{th}$ and $\theta^{c}_{th}$). If this is satisfied, then we can say that there is a link that a UAV can move.
\vspace{-3pt}
\begin{equation}
\begin{split}
& \mathit{L}^{\hat{p}}_p~\rightarrow~\mathit{D}^{\hat{p}}_p \wedge (\delta^{p',\hat{p},p}_{\alpha} \le \theta^{t}_{th}) \wedge (\delta^{p',\hat{p},p}_{\beta} \le \theta^{c}_{th})
\end{split}
\end{equation}

Similarly, if a UAV $u$ moves from $p'$ to $\hat{p}$ followed by $p$, then the deviation of the turn angle and climb angle to move from the path $p'$ to $\hat{p}$ and the path $\hat{p}$ to $p$ should be within user-defined thresholds.
\begin{equation}
\begin{split}
& \mathit{M}^{p', \hat{p}}_u \wedge \mathit{M}^{\hat{p}, p}_u~\rightarrow~(\delta^{p',\hat{p},p}_{\alpha} \le \theta^{t}_{th}) \wedge (\delta^{p',\hat{p},p}_{\beta} \le \theta^{c}_{th})
\end{split}
\end{equation}

A UAV visits a particular point from one of its neighboring points. If UAV $u$ has visited the point $p$, that means it has traveled from one of the neighboring points, $\hat{p}$.
\vspace{-3pt}
\begin{equation}
\begin{split}
& \mathit{V}^{p}_u~\rightarrow~ \bigvee_{\hat{p}}\mathit{M}^{\hat{p}, p}_u
\end{split}
\end{equation}

The starting point $S$ is the first point on the trajectory of any UAV. If the current point that a UAV $u$ is visiting is the starting point, then it has not come from any other point $\hat{p}$. This can be presented by the following equation:
\vspace{-3pt}
\begin{equation}
\begin{split}
& \forall_{\hat{p}} (p = S)~\rightarrow~\neg\mathit{M}^{\hat{p}, p}_u
\end{split}
\end{equation}

If a UAV $u$ has reached the destination point $D$, it should not move to anywhere from there. This is represented by the following equation:
\begin{equation}
\begin{split}
& \forall_{p} (\hat{p} = D)~\rightarrow~\neg\mathit{M}^{\hat{p}, p}_u
\end{split}
\end{equation}

A UAV can travel to a certain point only from another point (not from multiple points). If a UAV $u$ has moved from $\hat{p}$ to $p$, then it should not have come from any other point $\hat{p}'$.
\begin{equation}
\begin{split}
& \mathit{M}^{\hat{p}, p}_u~\rightarrow~\bigwedge_{\hat{p}' \neq \hat{p}}\neg\mathit{M}^{\hat{p}',p}_u
\end{split}
\end{equation}

Similarly, from any particular point, a UAV can only travel to one point and not multiple points. If UAV $u$ goes from $\hat{p}$ to $p$, then it must not have gone to any other point $p'$ from $\hat{p}$. This is formally represented as the following:
\begin{equation}
\begin{split}
& \mathit{M}^{\hat{p}, p}_u~\rightarrow~\bigwedge_{p' \neq p}\neg\mathit{M}^{\hat{p},p'}_u
\end{split}
\end{equation}

A UAV $u$ must visit the starting point $S$ and the destination point $D$. This is true for all the collaborating UAVs in the network and is presented by the following:
\begin{equation}
\begin{split}
& \forall_{u}(\mathit{V}^{S}_{u} \wedge \mathit{V}^{D}_{u})
\end{split}
\end{equation}

Some of the points where the UAVs can go are in close proximity to the data collecting sensors. These points need to be visited by the UAVs so that they can collect the data. If $\mathbb{S}$ is the set of all the points that have data-points, then at least a threshold ($\mathit{v}_{th}$) fraction of all the data-points must be visited by the UAVs during their journeys towards the destination.
\begin{equation}
\begin{split}
& \frac {\sum_{u,p \in \mathbb{S}}\mathit{V}^{p}_u}{\vert \mathbb{S} \vert}~\geq~\mathit{v}_{th}
\end{split}
\end{equation}

Some of the points in the three-dimensional space may have to be avoided by the UAVs while making their journeys, for these points may currently reside in restricted zones. If $\mathbb{F}$ denotes the set of all the forbidden points that should be avoided, then no UAV should visit them.
\begin{equation}
\begin{split}
& \bigwedge_{p \in \mathbb{F}}\neg\mathit{V}^{p}_u
\end{split}
\end{equation}

\subsection{Modeling UAV Fuel Consumption and Cost}
While calculating the fuel consumption, we not only consider the distance but also consider the turn and climb angles for going from a certain point to another. The direction of the turn angle has no effect on fuel consumption, but the direction of the climb angle does affect the amount of fuel consumption. If a UAV is going up against gravity, then it has to burn more fuel. On the other hand, if it is going down, then less fuel is required. We provision a positive climb angle if the UAV is going up with respect to its current direction, while it is counted as negative if it is doing downwards.
If $\mathit{d}^{\hat{p}}_{p}$ is the distance between point $\hat{p}$ and $p$, $\alpha^{\hat{p},p}_{u}$ denotes the turn angle of UAV $u$ for going to $p$ from $\hat{p}$, $\beta^{\hat{p},p}_{u}$ is the climb angle to move, and $\mathit{r}_{u}$ is the fuel consumption rate ({\it e.g.}, mpg) of a UAV $u$, then the required fuel, $\mathit{f}^{\hat{p},p}_{u}$, is calculated as follows:
\begin{equation*}
\begin{split}
& \mathit{f}^{\hat{p},p}_{u}~=~\frac{\mathit{d}^{\hat{p}}_{p} + k_1 \times \vert\alpha^{\hat{p},p}_{u}\vert + k_2 \times \beta^{\hat{p},p}_{u}}{\mathit{r}_{u}}
\end{split}
\end{equation*}
Here, $k_1$ and $k_2$ are simple constants that regulate the impact of turn and climb angles, respectively, of the UAV. The fuel usage does not depend on the direction of the turn angle $\alpha^{\hat{p},p}_{u}$, while it directly depends on the direction of the climb angle $\beta^{\hat{p},p}_{u}$. If the UAV is moving to a point located at a lower altitude than the source point, then it is traveling with gravity. As a result, there is less usage of fuel. On the other hand, if it has to travel against gravity, then fuel usage increases.

If $\mathit{c}_{e}$ is the cost of per unit fuel, then the cost, $\mathit{c}^{\hat{p},p}_{u}$, for a UAV for moving from $\hat{p}$ to $p$ is calculated as follows:
\begin{equation*}
\begin{split}
& \mathit{c}^{\hat{p},p}_{u}~=~\mathit{f}^{\hat{p},p}_{u} \times \mathit{c}_{e}
\end{split}
\end{equation*}

The cumulative cost for traveling to point $p$ for UAV $u$ is the cost up to the previous point $\hat{p}$, plus the cost to move from $\hat{p}$ to $p$, $\mathit{c}^{\hat{p},p}_{u}$.
\begin{equation}
\begin{split}
& \mathit{M}^{\hat{p},p}_u~\rightarrow~\mathit{C}^{p}_{u} = \mathit{C}^{\hat{p}}_u + \mathit{c}^{\hat{p},p}_{u}
\end{split}
\end{equation}

If $S$ is the starting point, the cost at the starting point for any UAV $u$ should be zero.
\begin{equation}
\begin{split}
& \mathit{C}^{S}_{u} = 0
\end{split}
\end{equation}

\subsection{Modeling UAV Travel Time}
We assume a constant speed for each UAV while it is making its flight from a point to another particular point. $\mathit{t}^{\hat{p},p}_{u}$ is the time taken by UAV $u$ to move from $\hat{p}$ to $p$ if $\mathit{s}_{u}$ is the speed of the UAV. This is calculated as follows:
\begin{equation*}
\begin{split}
& \mathit{t}^{\hat{p},p}_{u}~=~\frac{\mathit{d}^{\hat{p}}_{p}}{\mathit{s}_{u}}
\end{split}
\end{equation*}

$\mathit{T}^{p}_{u}$ is the time taken by UAV $u$ to reach up to $p$ from the starting point. $\mathit{H}^{\hat{p}}_{u}$ is the time of hovering of UAV $u$ at a previous point $\hat{p}$. We consider this hovering time to be either $0$ or $1$, for simplicity. A UAV may or may not hover over a point depending on the requirement of avoiding collision.
\begin{equation}
\begin{split}
& \mathit{M}^{\hat{p},p}_u~\rightarrow~\mathit{T}^{p}_{u} = \mathit{T}^{\hat{p}}_{u} + \mathit{t}^{\hat{p},p}_{u} + \mathit{H}^{\hat{p}}_{u}
\end{split}
\end{equation}
\begin{equation}
\begin{split}
& (\mathit{H}^{\hat{p}}_{u} = 0) \vee (\mathit{H}^{\hat{p}}_{u} = 1)
\end{split}
\end{equation}

If $\mathit{t}_{st}$ is the start time, then it is the time for all the UAVs at the starting point $S$.
\begin{equation}
\begin{split}
& \mathit{T}^{S}_{u} = \mathit{t}_{st}
\end{split}
\end{equation}

%

%

The following constraint ensures that no two UAVs visit the same point at the same time. This restriction is required to avoid collision between any two UAVs.
\begin{equation}
\begin{split}
& \mathit{V}^{p}_{u} \wedge \mathit{V}^{p}_{u'} ~\rightarrow~\mathit{T}^{p}_{u} \neq \mathit{T}^{p}_{u'}
\end{split}
\end{equation}

In fact, there should be a threshold time difference between the times of visit of any two UAVs to a certain point. If both UAVs $u$ and $u'$ visit the same point $p$, then there must be some time difference between their visits to avoid mid-air collision.
\begin{equation}
\begin{split}
& \mathit{V}^{p}_{u} \wedge \mathit{V}^{p}_{u'} ~\rightarrow~\mathit{T}^{p}_{u} - \mathit{T}^{p}_{u'} \geq \hat{t}_{th}
\end{split}
\end{equation}

\subsection{Modeling User Requirements}
One of the objectives of the trajectory planning model is to keep the budget within a limit. We consider the budget or cost limit for each UAV to be $\mathit{b}_{c}$, while $\mathit{b}_{t}$ is the budgeted time or time boundary within which the flights need to be performed.
The cost of the fuel by UAV $u$ should be less than a user-specified budget, $\mathit{b}_{c}$.
\begin{equation}
\begin{split}
& \mathit{C}^{D}_{u} \leq \mathit{b}_{c}
\end{split}
\end{equation}

Each UAV should reach the destination point from the starting point within a certain time limit, $\mathit{b}_{t}$.
\begin{equation}
\begin{split}
& \mathit{T}^{D}_{u} \leq \mathit{b}_{t}
\end{split}
\end{equation}

Again, a point cannot be reached by the UAVs by taking more cost than the budget or the time allows. If a UAV visits point $p$, then the cost for reaching up to that point should be less than the budget.
\begin{equation}
\begin{split}
& \mathit{V}^{p}_{u}~\rightarrow~\mathit{C}^{p}_{u} \leq \mathit{b}_{c}
\end{split}
\end{equation}

The time taken in reaching any point $p$ should be within the time limit.
\begin{equation}
\begin{split}
& \mathit{V}^{p}_{u}~\rightarrow~\mathit{T}^{p}_{u} \leq \mathit{b}_{t}
\end{split}
\end{equation}

\subsection{Modeling Resiliency Specifications}

Our trajectory modeling considers resiliency with respect to attack/contingency scenarios. The satisfaction of the resiliency specification provides  resilient trajectory plans. The resiliency specification is stated such that at least a certain percentage ($r_{th}$\%) of the surveillance points are (securely) covered by a set of UAVs (Equation 27), even if a certain number ($r$) of UAVs among them fail to operate or are compromised. 

We abstract the resiliency requirement by introducing constraints on surveillance requirements. If $p$ is a data point ($p \in \mathbb{S}$) and it satisfies the $r-$resiliency data collection, then at least $r + 1$ number of UAVs must visit (collect data at) this point during their travel to the destination. That is, to provide a single ($r = 1$) UAV failure resiliency to a surveillance point, two ($r + 1$) UAVs will need to visit the point. If $R^p_c$ specifies that if point $p$ is under $r-$resilient surveillance, we can formalize the above resiliency requirement as follows:
\begin{equation}
\forall_{p \in \mathbb{S}}~R^p_c \rightarrow~ \sum_{u} V^p_u \ge r + 1
\end{equation}

The resilient trajectory design often needs to satisfy further requirements. A requirement can be data freshness or closeness constraint that ensures that the times at which data is collected by $k + 1$ UAVs at a data point should not be apart from one another more than a threshold value (say $t_\mathit{th}$), so that the collected data records can be verified for any discrepancy. We call such a data point to be under $k-$resilient surveillance. Such a constraint may differ for different data points based on the data change frequency (which may depend on the data type, location, etc.) and/or the criticality of the data. This constraint ensures that even if $k$ UAVs fail to visit a point $p$, there is at least a UAV $u$, which along with another UAV $u'$ will cover this point. This constraint is formalized below:
\begin{equation}
\forall_{p \in \mathbb{S}}~R^p_t \rightarrow~ \bigvee_{u} \sum_{u', u \neq u'} \hat{V}^p_{u, u'} \ge k + 1
\end{equation}
Here, $\hat{V}^p_{u, u'}$ specifies that if UAVs $u$ and $u'$ visit point $p$, then the visiting times ($u$ earlier than $u'$, required for ordering with respect to the visiting time) are no more than the threshold time apart. This is defined as follows:
\begin{equation}
\hat{V}^p_{u, u'} \rightarrow~ V^p_u \wedge V^p_u \wedge (T^p_u \le T^p_{u'}) \wedge (T^p_{u'} - T^p_{u} \le {t}^{p}_\mathit{th})
\end{equation}

We need to make sure that at least a certain percentage ($\mathit{k_{th}}$\%) of the data points are covered by the UAVs such that the points follow the $k-$resiliency property.
%
%
\begin{equation}
(\frac{\sum_{p}R^p_c}{\vert \mathbb{S}\vert} \ge \mathit{r_{th}}) \wedge (\frac{\sum_{p}R^p_t}{\vert \mathbb{S}\vert} \ge \mathit{k_{th}})
\end{equation}
\section{Implementation and A Case Study}
\label{Sec:Imp}
In this section, we present the implementation details of our proposed model. We also present a detailed case study for further explaining how the model works. The main objective of our configuration synthesis problem is to synthesize the trajectories of the UAVs by satisfying various requirements, as well as the user's business and operating constraints. The synthesis problem is formalized as the satisfaction of the conjunction of all the constraints in the equations of Section~\ref{Sec:Model}.

\subsection{SMT Encoding and Target Variables}
\label{SubSec:encode}
 We implement Synth4UAV by encoding the above formal model into SMT formulas~\cite{Moura09}. In this encoding purpose, we use Z3, an efficient SMT solver~\cite{de2008z3}.
The solver checks the verification constraints and provides a satisfiable (SAT) result if all the constraints are satisfied. The SAT result provides a SAT instance, which represents the value assignments to the parameters of the model. According to our objective, we are mainly interested in the assignments to the following variables: (i) the decision variable referring to whether a UAV $u$ has visited a particular point $p$, $\mathit{V}^{p}_u$ and (ii) the variable denoting whether a UAV $u$ travels from a point $\hat{p}$ to $p$, $\mathit{M}^{\hat{p}, p}_u$. We can retrieve the trajectories for all the UAVs with the values of these two variables. The assignments of the variables are printed in an output text file. The values of other variables such as $\mathit{T}^{p}_{u}$ can provide information on which point was traveled at what time, while $\mathit{C}^{p}_{u}$ can tell about the fuel cost associated with traveling up to certain points.
Appendix~\ref{App_Code} presents an excerpt from the generated z3 code. It shows a few of the expressions that assert the constraints according to our trajectory synthesis model.

Throughout the model of the network synthesis, for simplicity, we have used the sum of Boolean variables several times for formally modeling the constraints. While encoding the summation of Boolean variables, we have created corresponding integer variables ($0$ or $1$) and used their sum for calculation.

\subsection{Optimal Trajectory Synthesis}
\label{SubSec:Optimal}

\begin{algorithm}
\small
\caption{Optimal Budget Determination}
\label{Algo_Optimal}
\begin{algorithmic}[1]

\State $\mathit{b}_{c}^{min} \coloneqq \mathit{0}$
\State $\mathit{b}_{c}^{max} \coloneqq \mathit{b}_{c}$

\If{Solver returns SAT}
	\State Get Model, $\mathit{M}$
	\State $\mathit{b}_{c}^{max} \coloneqq \mathit{C}_{u}^{d}$
	\State $C = 0$

	\Repeat
		\State $\mathit{b}_{c} \coloneqq (\mathit{b}_{c}^{min} + \mathit{b}_{b}^{max}$) / 2
		\State Update the constraints associated to $\mathit{b}_{c}$
		\If{Solver returns SAT}
			\State Get Model, $\mathit{M}$ and Get updated $\mathit{C}_{u}^{d}$
			\State Update $\mathit{b}_{c}^{max}$ with new value of $\mathit{C}_{u}^{d}$
		\Else
			\State
			$\mathit{b}_{c}^{min} \coloneqq \mathit{b}_{c}$
		\EndIf
		\State
		$\mathit{C} \coloneqq \mathit{C} + 1$
	\Until{($\mathit{b}_{c}^{max} - \mathit{b}_{c}^{min} \approx 0) \parallel (C = \mathit{C}^{max}$)}
\EndIf

\end{algorithmic}
\end{algorithm}

The synthesis result represents a comprehensive trajectory plan for the UAVs to perform the intended tasks. There is usually more than one satisfiable model that can satisfy the constraints. 
These models will provide different costs and times for the overall process, although all of them will be less than the budgeted cost ($\mathit{b}_{c}$) and time ($\mathit{b}_{t}$).
We can choose the most cost- and time-efficient trajectory plans among all alternative satisfiable models for the same set of constraints.
We use Algorithm~\ref{Algo_Optimal} which updates the minimum ($\mathit{b}_{c}^{min}$) and maximum ($\mathit{b}_{c}^{max}$) values of the budget, and finds the optimal trajectory plan. Their values are used to update $\mathit{b}_{c}$, which is used to try for a more optimal solution. Initially, $\mathit{b}_{c}^{max}$ is set to the initial user-defined budget ($\mathit{b}_{c}$) and $\mathit{b}_{c}^{min}$ is set to zero. We use a binary search-based method to find the optimal budget.

However, finding the optimal solution using the algorithm usually requires more time than a single satisfiable model. The algorithm requires several invocations of the model solver and often confronts UNSAT results, which usually take a longer time. The time complexity of this algorithm becomes $\mathcal{O}(T \times \log_2 D)$, where $T$ is the time for one model solution and $D$ is the difference between $\mathit{b}_{c}^{min}$ and $\mathit{b}_{c}^{max}$.
The user can control the number of iterations of the model ($\mathit{C}^{max}$) and generate a sub-optimal solution if required.

\subsection{Trajectory Smoothing}
\label{Subsec:smooth}
The synthesized trajectories can be smoothed by applying a 3D Bezier curve interpolation~\cite{sahingoz2014generation}. Careful studies need to be done to make sure that the smooth trajectories remain adequately close to the data sources so that the UAVs and the sources are within range of each other. While this is out of the scope of this research, we leave it for our future works.

\subsection{An Example Case Study}
\label{Subsec:case}

UAVs can be used to collect data from sensors deployed in remote points, which is essentially an alternative to the use of UAVs in a surveillance purpose. This is efficient technology when it is not feasible (either technically or economically) to deploy necessary communication infrastructures. If sensors cannot transmit data by themselves to the data center due to the lack of proper infrastructure or energy-saving purposes, UAVs are convenient options to collect data from them. We need to make sure that a certain percentage of the points with sensors are covered by the set of UAVs, even if a certain number of UAVs fail while in flight.

In this case study, we consider 30 waypoints in the three-dimensional coordinate space, of which 15 contain data sources. The coordinates are depicted in the input file presented in Appendix~\ref{App_input_output}. For example, (100, 100, 0), (500, 1000, 100), (500, 2000, 100), etc. are some of the points that the UAVs can travel to. In other words, these points may or may not be parts of the trajectories of the UAVs. The indices of the data points are provided from the input file. The start point and end point of the UAV trajectories are specified, which can be the coordinates of base stations. In this case, point 1 is the starting point and point 18 is the ending point for all the UAVs. There are four forbidden points, 6, 14, 19, and 30, which should be avoided by the UAVs while flying from the starting point to the ending point. 

We consider 5 UAVs. The average velocity, mileage, and angle with $X$ axis (at the beginning) of the UAVs are also provided. The restrictions of turn and climb angles (in degrees) of the UAVs while moving from one point to another are given in the input file. We consider the deviation of the moving direction of the UAVs. That is, after a UAV reaches a particular point, how much it can deviate from its current trajectory both vertically and horizontally in order to move to the next point.

\vspace{3pt}
\textbf{A Satisfiable Case:}
The data coverage threshold is selected to be 80\% in this case study. That means, at least 80\% of the data points must be covered by the UAVs while they are making their journeys. The data freshness threshold specifies the time (in seconds) within which a certain number of UAVs must visit a data point to ensure the freshness of the data. At the beginning of this case study, we consider this to be 20 sec. At this point, we consider the value of $k$ to be 2, while the resilient data coverage is chosen to be 70\%. This means there should be at least 70\% of the data points, such that they are covered by at least $k = 2$ UAVs every 20 seconds. We also need to make sure that at least 70\% of the data points are covered by a minimum of $k$ UAVs.

The price of fuel is provided in the input file. The available budget and the total available time within which the UAVs need to finish their task by traveling from the start point to the end point are also provided. For the first run of the tool in this case study, we consider a budget of \$6000, and the total allowed time is chosen to be 2000 sec.

\begin{figure}
	\centering
	\includegraphics[keepaspectratio,scale=0.25,bb=0 0 1704 1029]{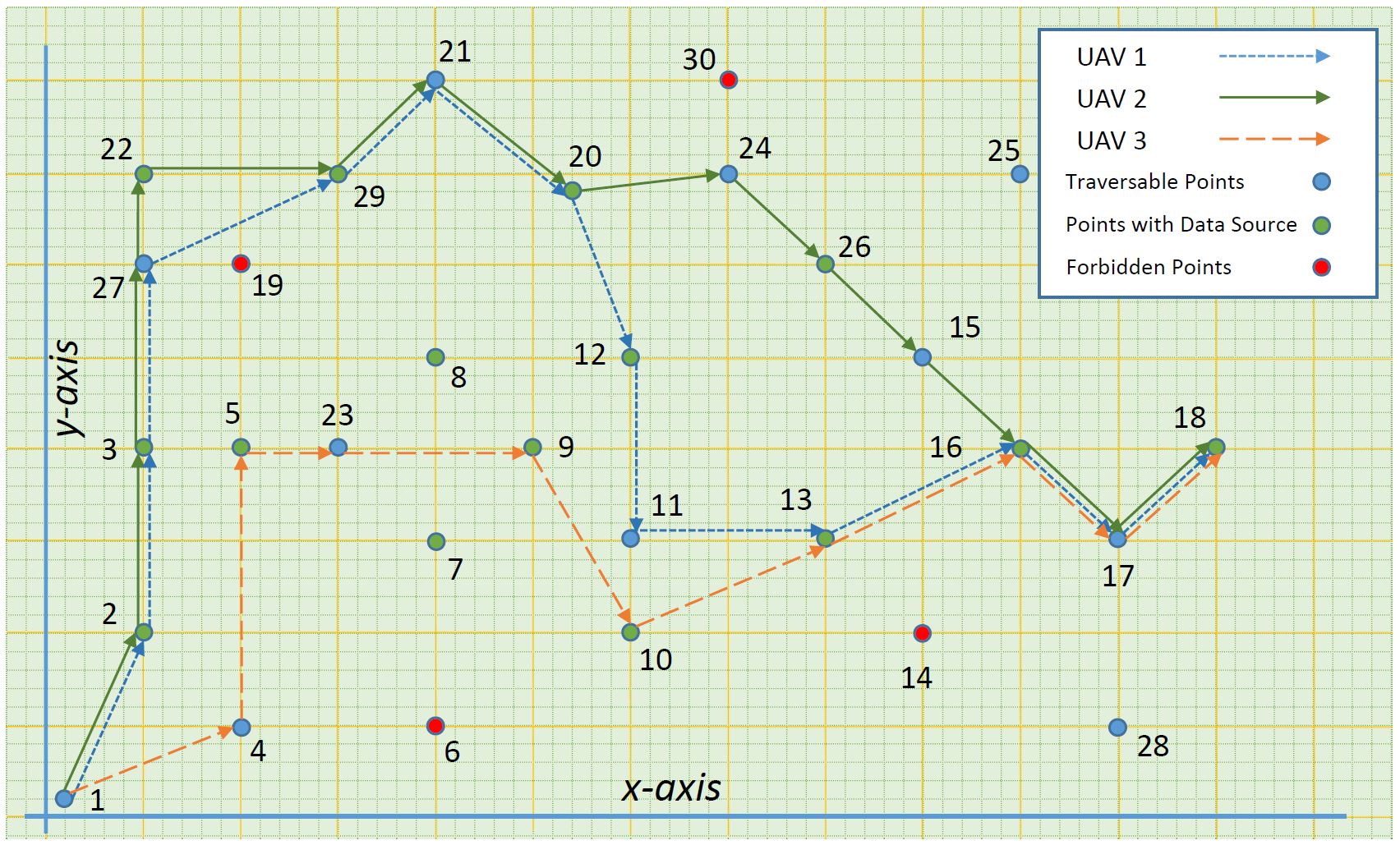}
	\caption{Generated trajectories of three of the five UAVs on $XY$ plane visiting waypoints near data sensors.}
	\label{Fig_CaseTraj}
\end{figure}

Now we try to solve this multi-objective problem in order to find the trajectories of the UAVs. The solution should provide the paths for each of the UAVs that satisfy all the conditions mentioned in Section~\ref{Sec:Model}.
In this case, the solver gives a SAT result. The result is printed in an output file, which is presented in Appendix~\ref{App_input_output}. The output file provides the times at which the UAVs visit each of the points on their trajectories. For example, UAV 1 is at point 1 at the beginning (0 sec), it is at point 2 at 20.79 sec, at point 3 at 41.79 sec, and so on. If a UAV needs to hover over any point to avoid a mid-air crash, the time of hovering is also presented in the output file. The output also provides the trajectories. For example, UAV 1 visits the following points during its journey in sequence: 1, 2, 3, 27, 29, 21, 20, 12, 11, 13, 16, 17, and 18. Similarly, the trajectories of all other UAVs are printed. Fig.~\ref{Fig_CaseTraj} presents a top view of the overall trajectory plan for 3 of the 5 UAVs on the $XY$ plane. The data points are colored green, while the forbidden ones are red. All other points are colored blue. All of the UAVs start from waypoint 1 and end their flight at waypoint 18. During their flights, they cover the points that have the data sources with the required resiliency.

\vspace{3pt}
\textbf{An Unsatisfiable Case:}
For the next run, we increase the value of $k$ to 3. In this scenario, the solver provides an UNSAT result, which means there is no satisfiable solution to the problem, and there cannot be any trajectory for the UAVs unless some conditions are modified. Increasing the budget, number of UAVs, etc. may provide us with a SAT result.
\section{Evaluation}
\label{Sec:Eval}

\begin{figure*}[t]
\begin{center}
\subfigure[]{
\label{Graph_BudgetVsCov}
\includegraphics[scale=0.43, keepaspectratio=true]{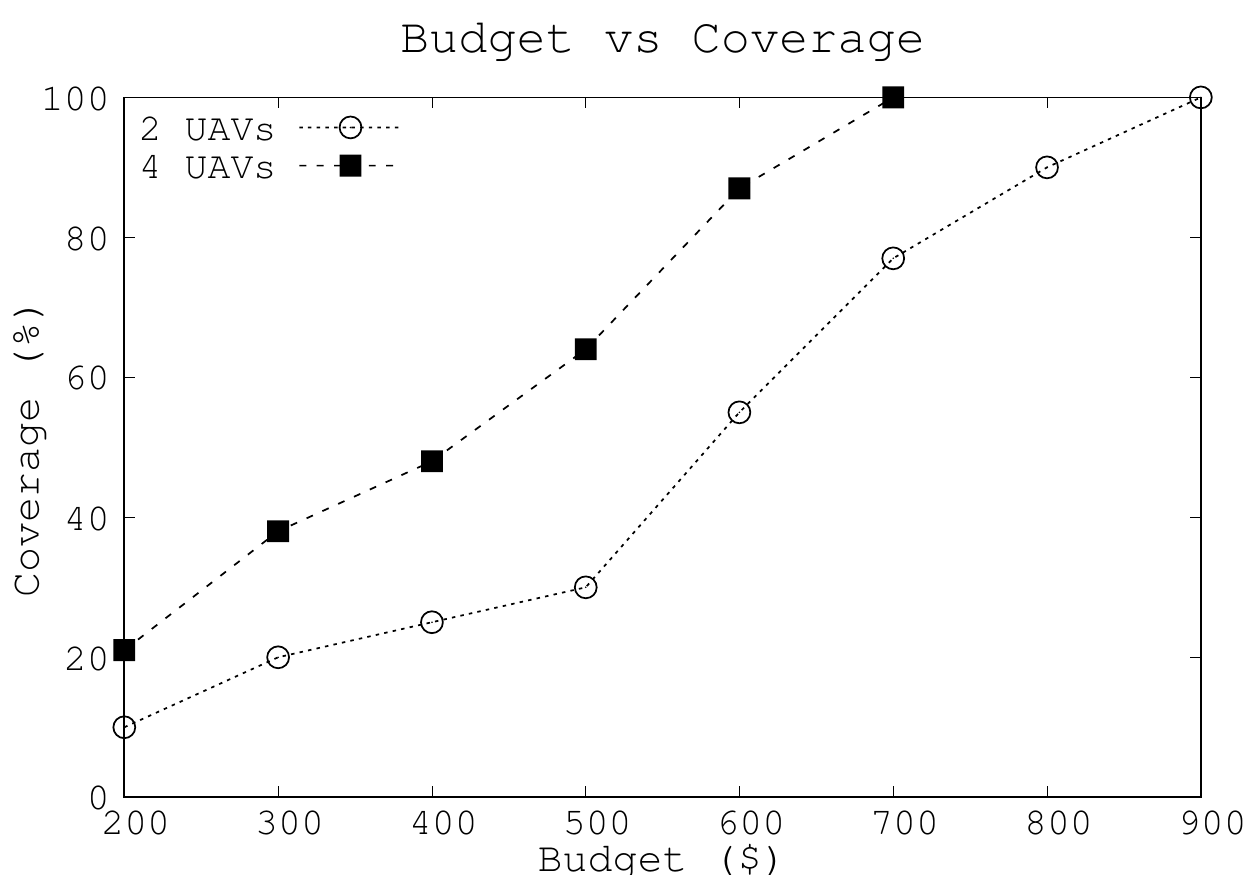}
}
\subfigure[]{
\label{Graph_UAVVsCovTime}
\includegraphics[scale=0.43, keepaspectratio=true]{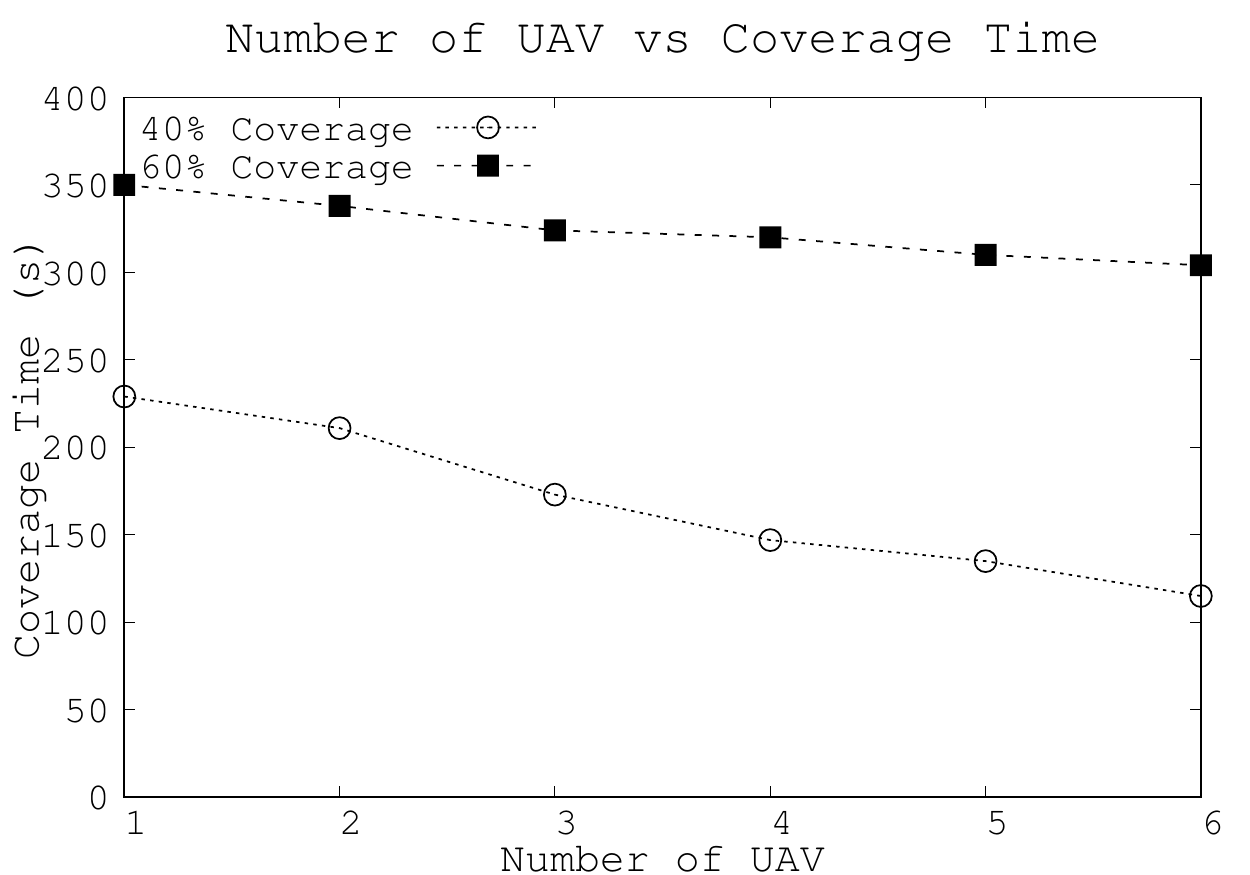}
}
\end{center}
\vspace{-10pt}
\caption{(a) Data points coverage w.r.t. budgeted cost, (b) Data points coverage time w.r.t. number of UAVs.}
\label{Graph_RelUAVTraj}
\end{figure*}

\begin{figure*}[t]
\begin{center}
\subfigure[]{
\label{Graph_CovVsUAV}
\includegraphics[scale=0.43, keepaspectratio=true]{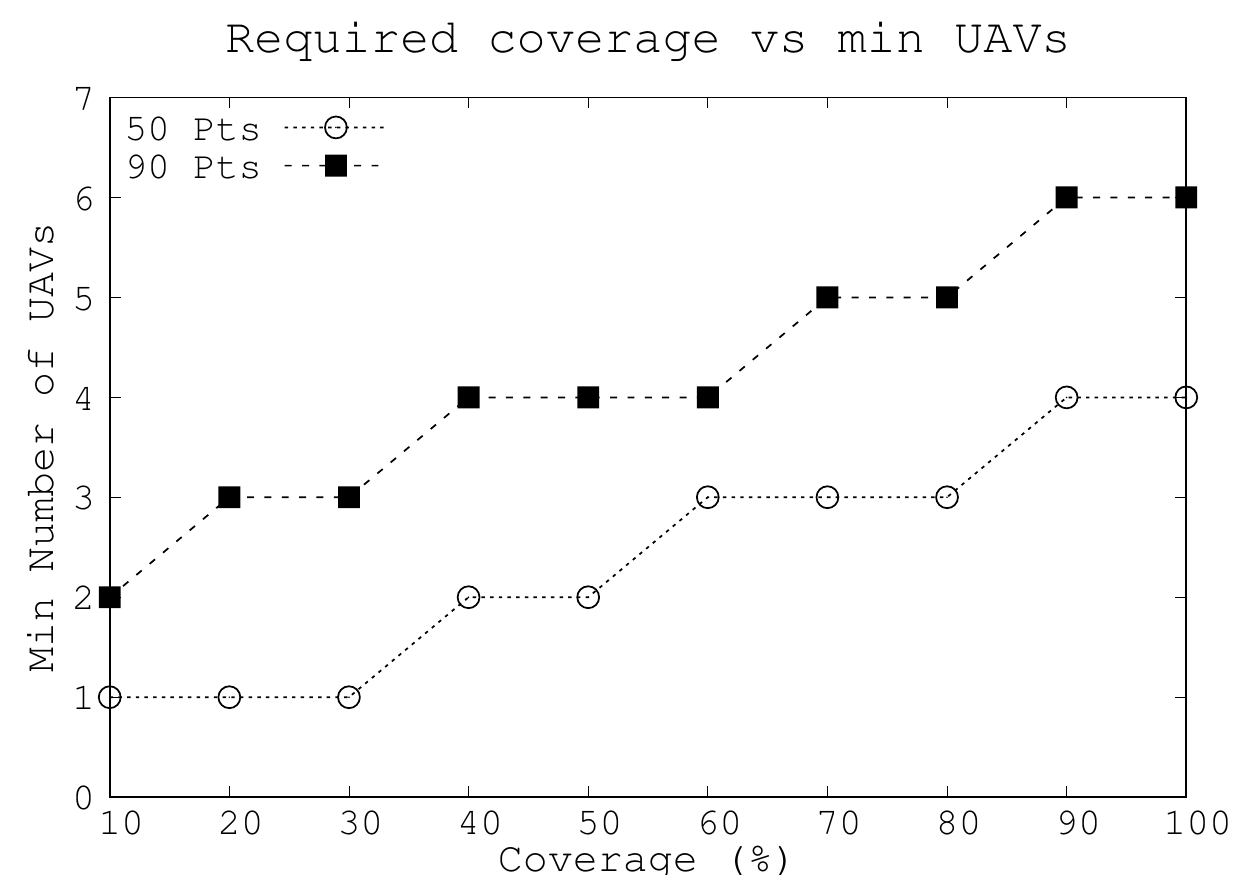}
}
\subfigure[]{
\label{Graph_BudgetTimeVsCov}
\includegraphics[scale=0.43, keepaspectratio=true]{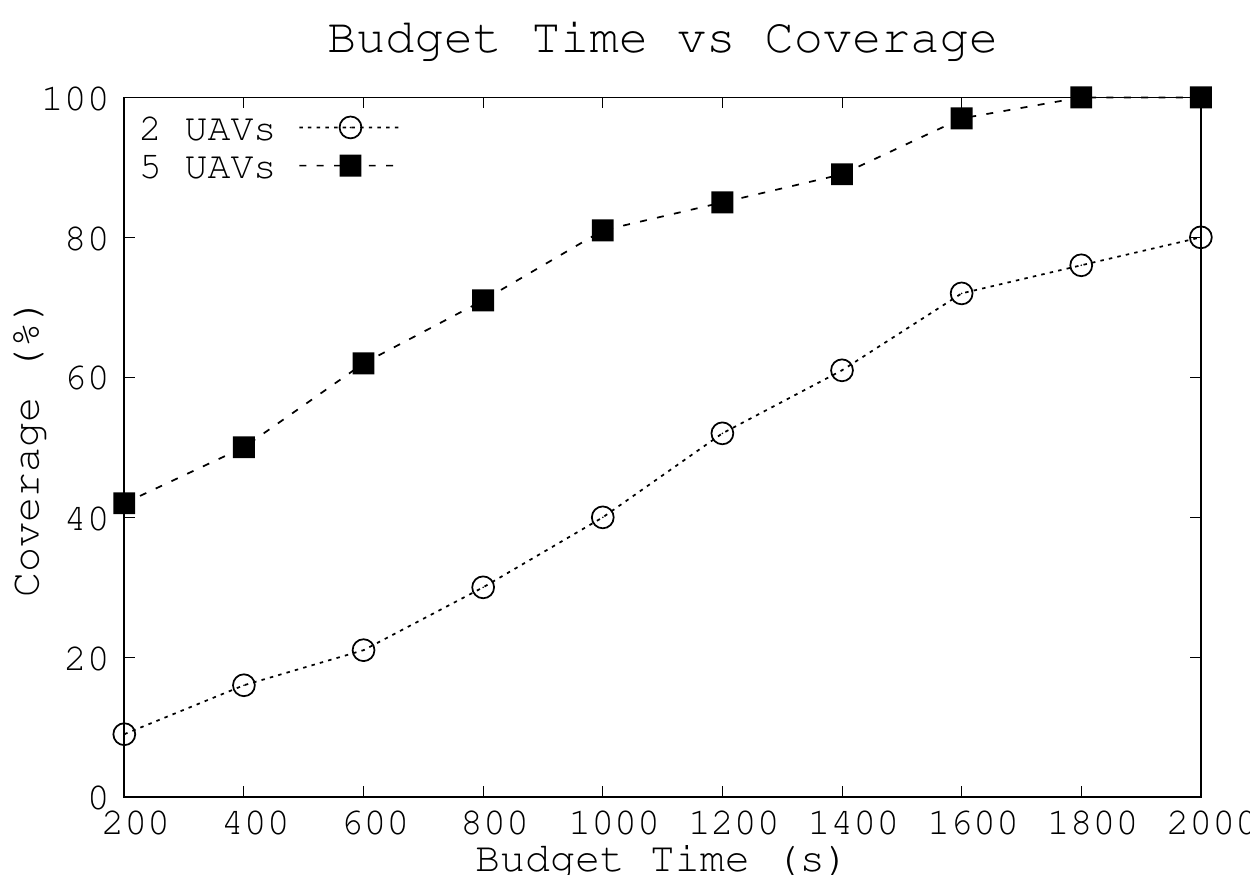}
}
\end{center}
\vspace{-10pt}
\caption{(a) Minimum number of required UAVs w.r.t. required percentage of coverage, (b) Data points coverage w.r.t. budgeted time.}
\label{Graph_RelUAVTraj}
\vspace{-15pt}
\end{figure*}

\begin{figure*}[t]
\begin{center}

\subfigure[]{
\label{Graph_DataPtVsBudget}
\includegraphics[scale=0.43, keepaspectratio=true]{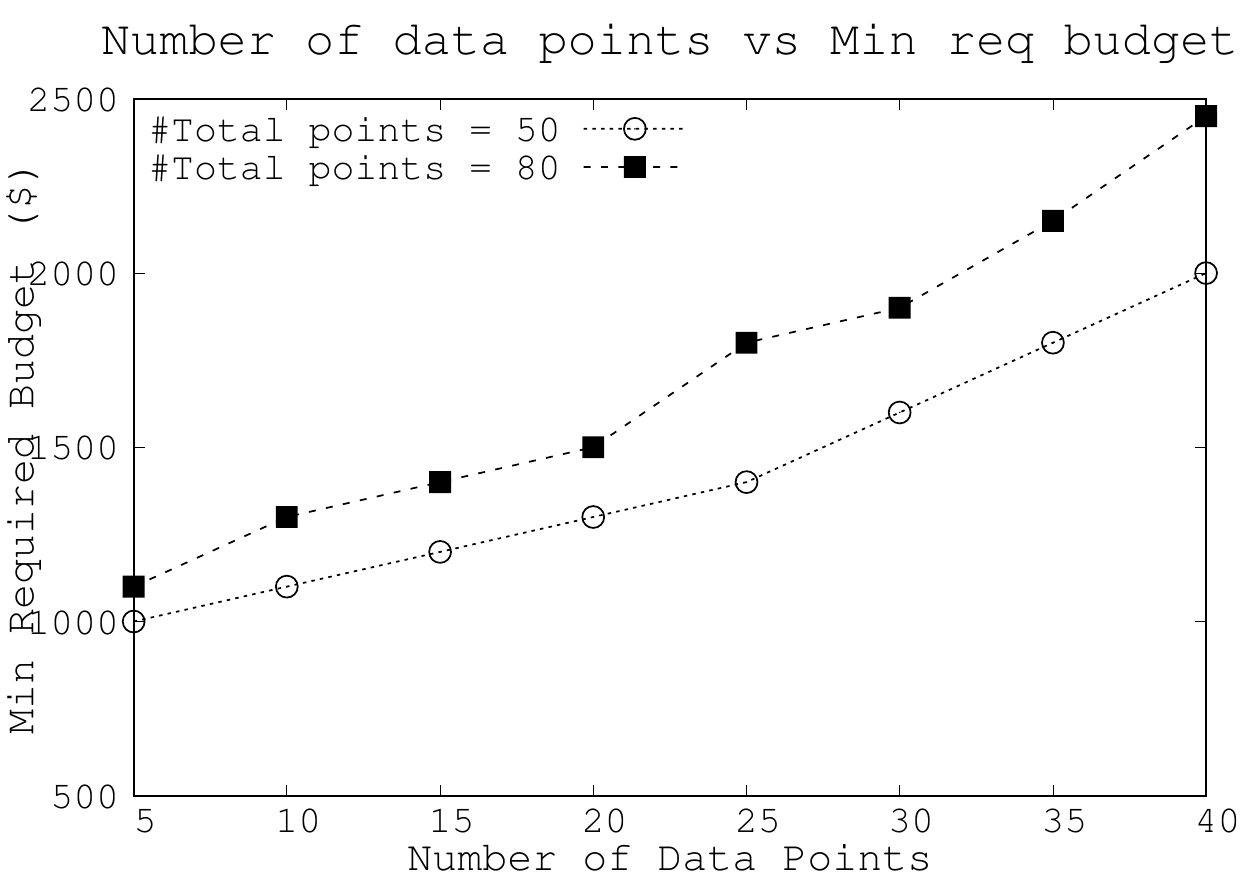}
}
\subfigure[]{
\label{Graph_kResVsBudget}
\includegraphics[scale=0.43, keepaspectratio=true]{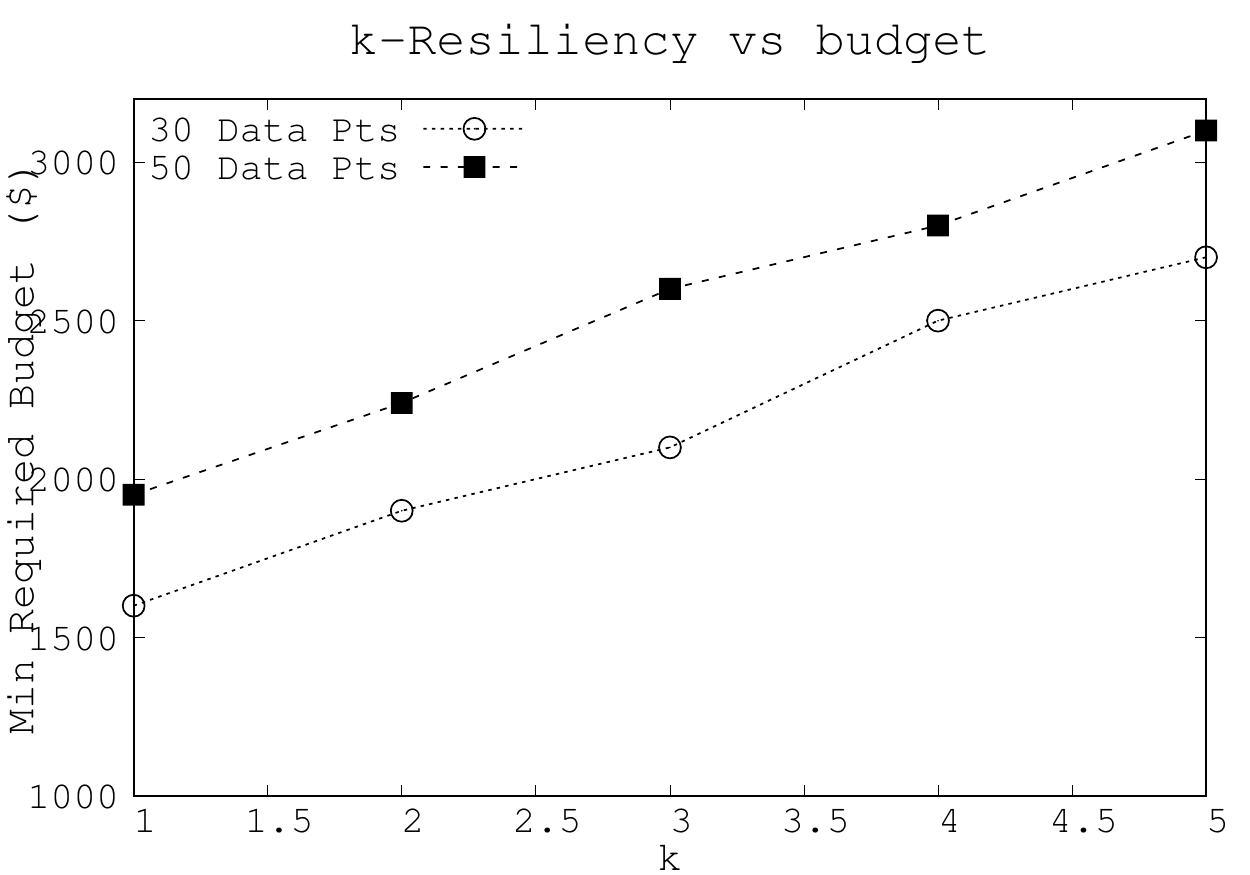}
}
\end{center}
\vspace{-10pt}
\caption{(a) Minimum required budget w.r.t. number of data points and (b) Minimum required budget w.r.t. $k-$ resiliency.}
\label{Graph_RelUAVTraj2}
\vspace{-10pt}
\end{figure*}

In this section, we present the evaluation of the proposed framework for synthesizing the UAV trajectories. 

\subsection{Methodology}
\label{subsec:method}
We create synthetic problem sets in a three-dimensional Cartesian space, where there are different points indicating the probable points for UAVs to fly. We randomly specify some data points (50-80\% of all the points), which must be covered by the UAVs while performing their flights from a pre-specified starting point to an end point. There are several forbidden points in the space as well, indicating the areas of geographical obstacles or unauthorized zones.

We run the tool that implements the formal model on a Windows 10 machine equipped with an Intel Core i5 processor and 12 GB of memory. We change different parameters of the model from the input file and observe the scalability of the tool, as well as the relationships between the parameters. The evaluation heavily depends on the placement of the available points in the 3D space and the allowed ranges of the UAVs. The maximum allowed turn angle and climb angle are chosen according to the placements of the points.

\subsection{Relationship between Different Parameters}
\label{subsec:rel}
We present the relationship among different parameters in Fig.~\ref{Graph_RelUAVTraj} and \ref{Graph_RelUAVTraj2}. We observe the maximum coverage produced by a SAT result while varying the budget for each UAV in Fig.~\ref{Graph_BudgetVsCov}. We present the analysis for two cases: one with 2 UAVs and the other with 4 UAVs while allowing adequate time for coverage. It is obvious that more points can be covered by the UAVs if there is more budget for fuel. If we use more UAVs, then we can cover 100\% of the data points with less budget per UAV. For example, if we use 2 UAVs, then we can cover 77\% of the data points with a budget of \$700 per UAV. On the other hand, using 4 UAVs with the same budget allows us to cover all the data points.

In Fig.~\ref{Graph_UAVVsCovTime}, we present the coverage time of the data points for the different numbers of UAVs. We show results for the requirement of 40\% coverage and 60\% coverage. If there are more than 1 UAVs, they reach the destination point at different times, as they may have different speeds and are not allowed to visit the same point at the same time. We take the time required by the last UAV to reach the destination. We can observe that the required time decreases slowly if we have more UAVs to cover our intended percentage of data points.

We present the minimum number of required UAVs to cover different percentages of the data points in Fig.~\ref{Graph_CovVsUAV}. As we increase our requirement of coverage, the minimum number of required UAVs also increases if we keep the budget cost per UAV, and the total budgeted time constant. We show the results for two problem sets: one with 50 points and another with 90 points.

The change in the percentage of coverage with respect to the budgeted time is presented in Fig.~\ref{Graph_BudgetTimeVsCov}. This is the maximum total time allowed for all the UAVs to reach the end point of their trajectories. We show two cases for 2 UAVs and 5 UAVs for a total of 100 points and 40 data points. We keep the values of $k$ and $r$ as 2 and 3, respectively. We can observe that as the allowed time increases, more of the data points can be covered. The higher number of UAVs means that more of the data points can be covered. For a budgeted time of 1800 sec, 5 UAVs can visit all the data points and reach the final point on their trajectories.

Next, we present the minimum required budget per UAV while varying the number of data points in Fig.~\ref{Graph_DataPtVsBudget}. We keep the total number of points, required coverage, the values of $k$ and $r$, etc. unchanged for this evaluation. The minimum required budget is the budget for which the solver returns a SAT answer, which means that trajectories are possible. As expected, the required budget increases with the increment of the number of data points.

The minimum required budget per UAV with the change of the value of $k$ is presented in Fig.~\ref{Graph_kResVsBudget}. We can observe that with the increment of $k$, the required budget increases quickly. As the value of $k$ increases, higher numbers of UAVs need to cover more data points, which results in the complex trajectories of the UAVs. Hence, a bigger budget is required to cover the fuel cost of the UAVs.
%

\begin{figure*}[t]
\begin{center}
\subfigure[]{
\label{Graph_PointVsTime}
\includegraphics[scale=0.43, keepaspectratio=true]{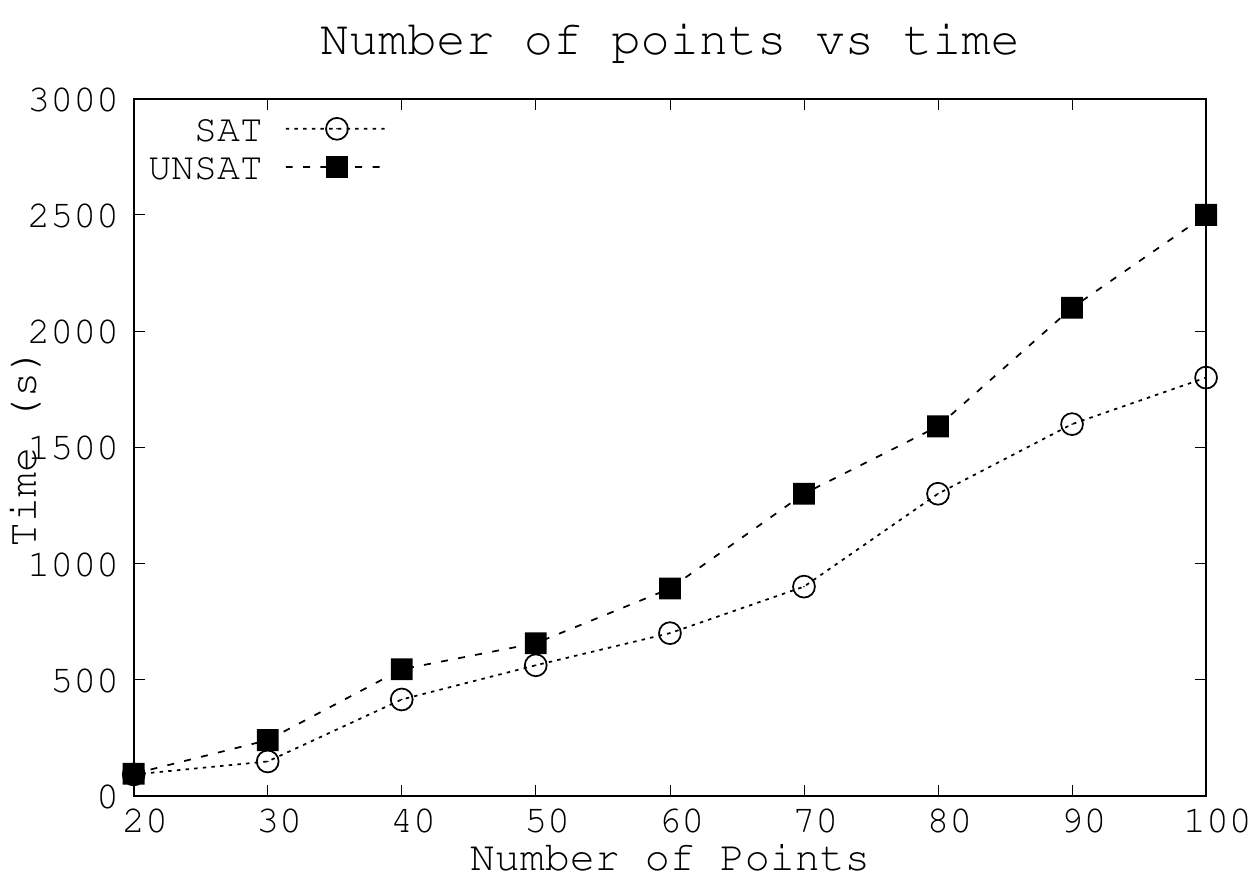}
}
\subfigure[]{
\label{Graph_UAVVsTime}
\includegraphics[scale=0.43, keepaspectratio=true]{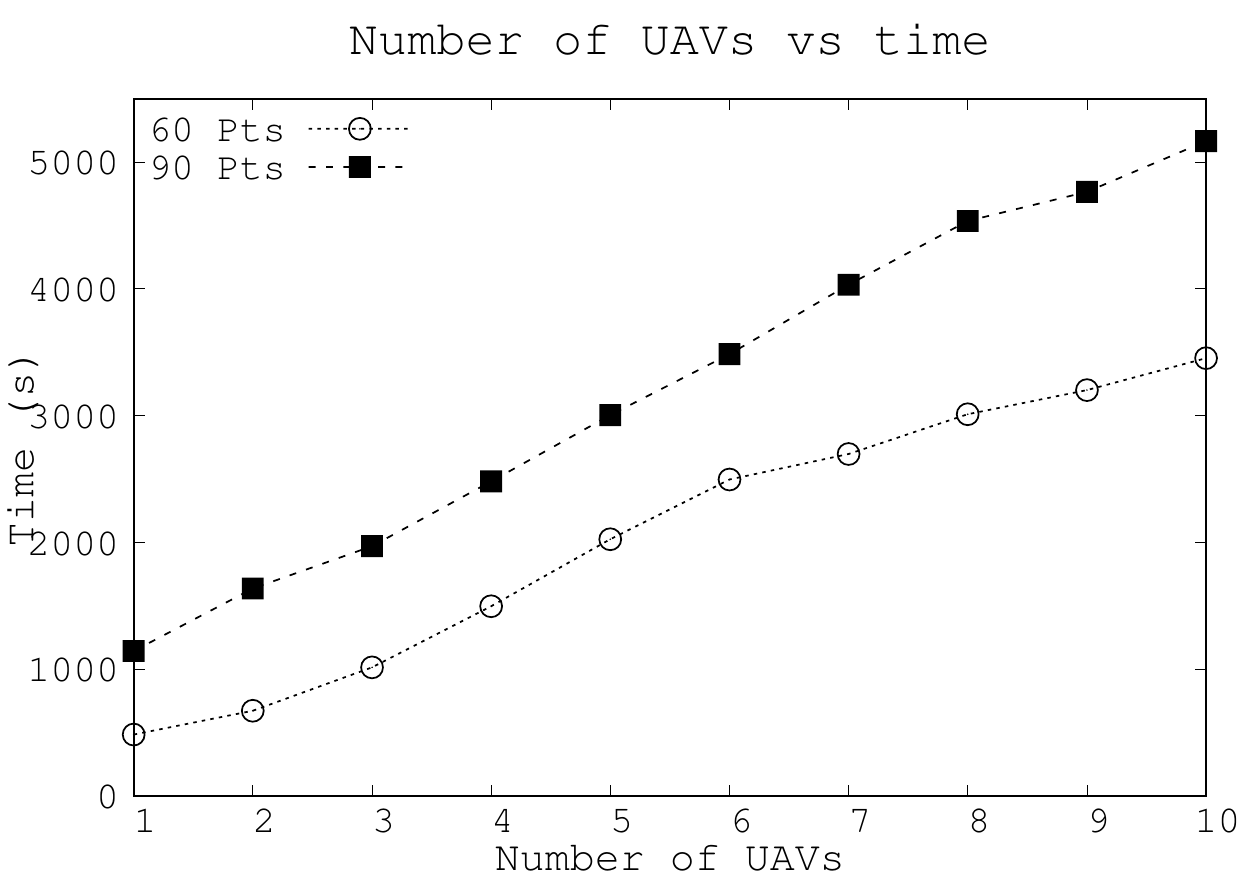}
}
\end{center}
\vspace{-10pt}
\caption{UAVTraj model synthesis time (a) w.r.t. number of points, (b) w.r.t. number of UAVs.}
\label{Graph_ScaleUAVTraj}
\end{figure*}
\begin{figure*}[t]
\begin{center}
\subfigure[]{
\label{Graph_CovVsTime}
\includegraphics[scale=0.43, keepaspectratio=true]{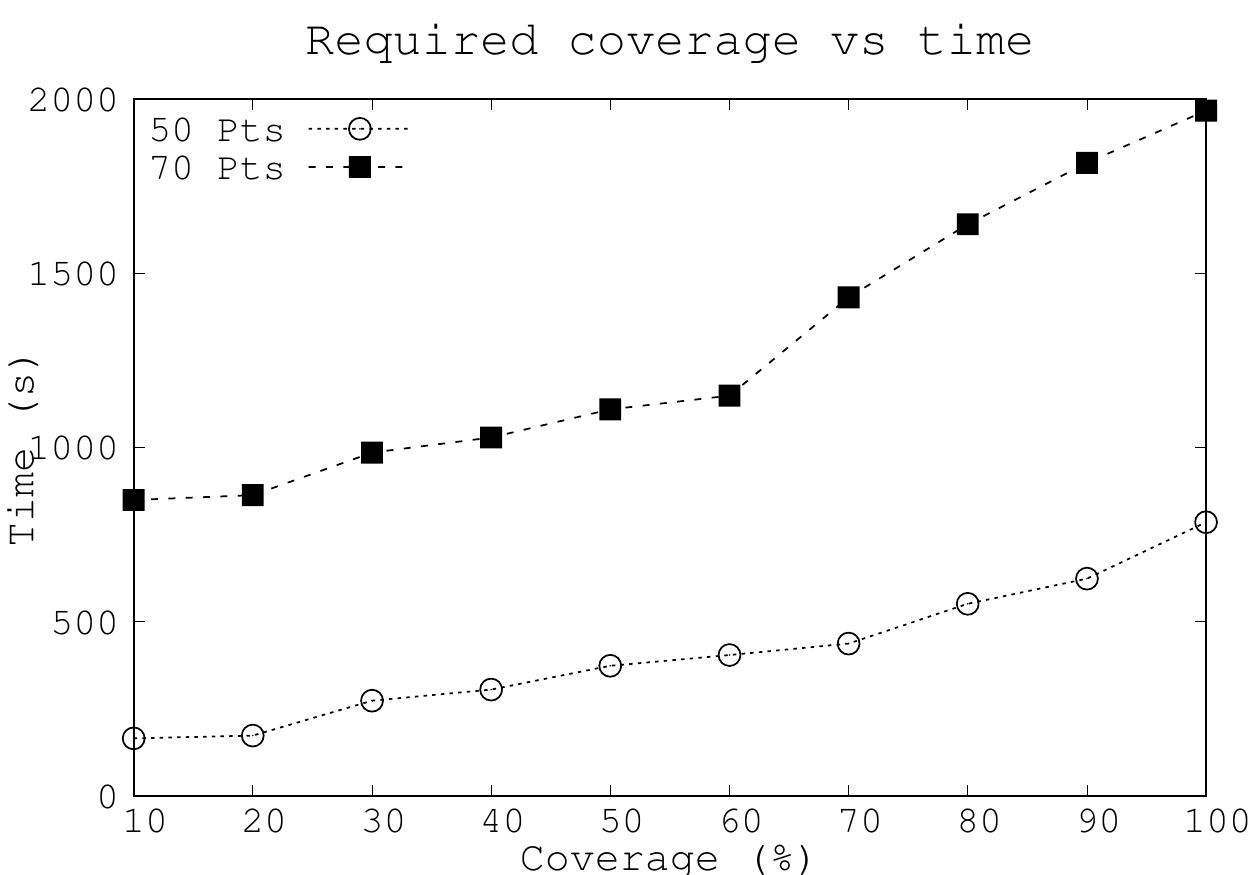}
}
\subfigure[]{
\label{Graph_DataPtVsTime}
\includegraphics[scale=0.43, keepaspectratio=true]{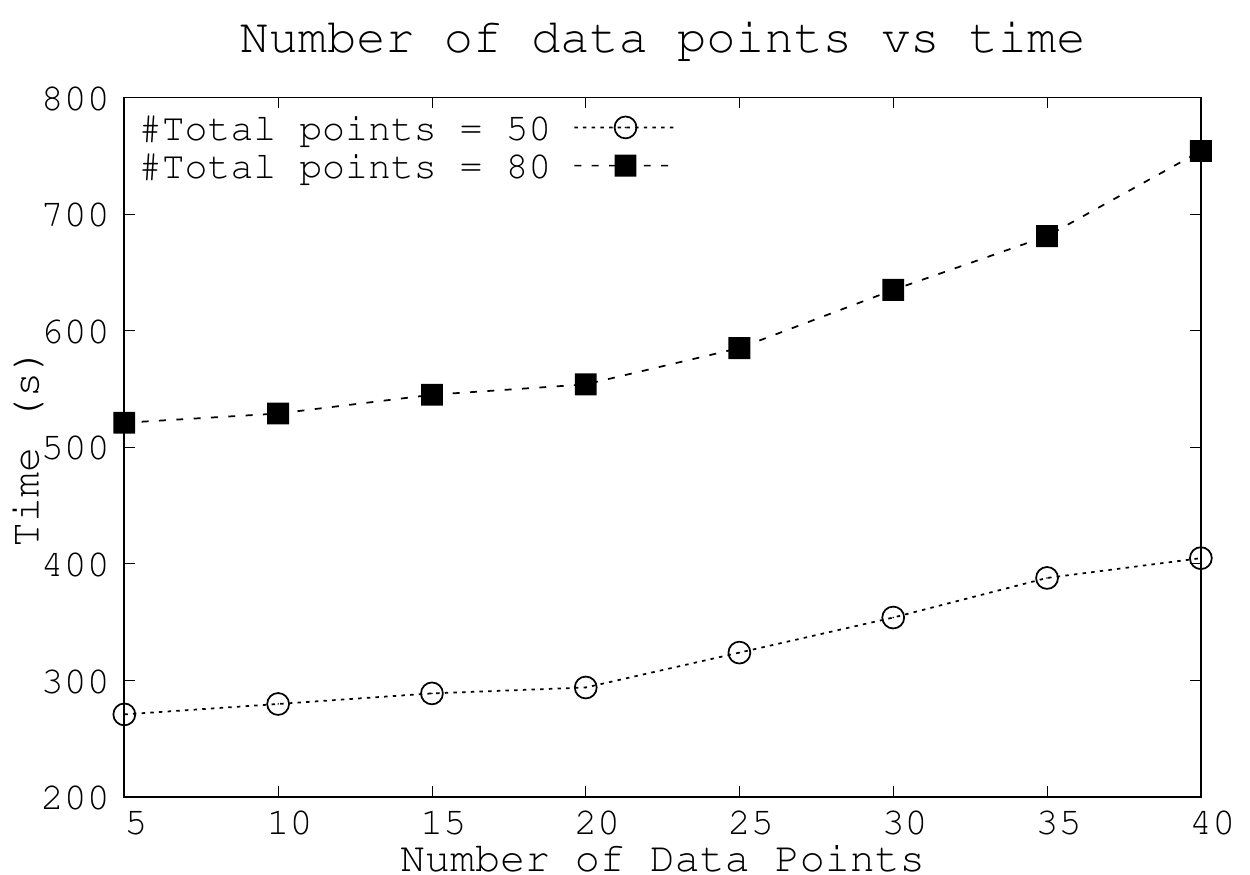}
}
\end{center}
\vspace{-10pt}
\caption{UAVTraj model synthesis time (a) w.r.t. required percentage of coverage, (b) w.r.t. number of data points.}
\label{Graph_ScaleUAVTraj2}
\end{figure*}
\begin{figure*}[t]
\begin{center}
\subfigure[]{
\label{Graph_rResVsTime}
\includegraphics[scale=0.43, keepaspectratio=true]{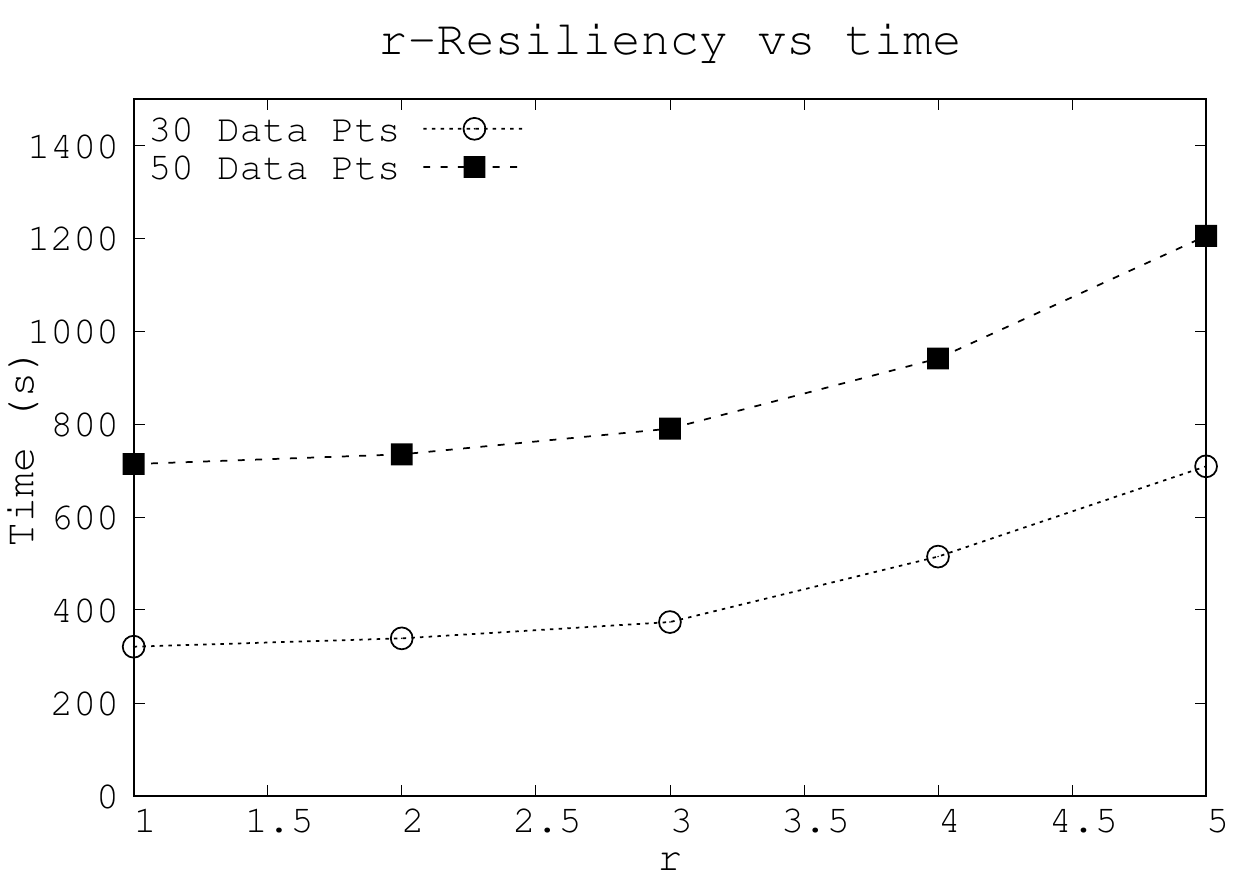}
}
\subfigure[]{
\label{Graph_kResVsTime}
\includegraphics[scale=0.43, keepaspectratio=true]{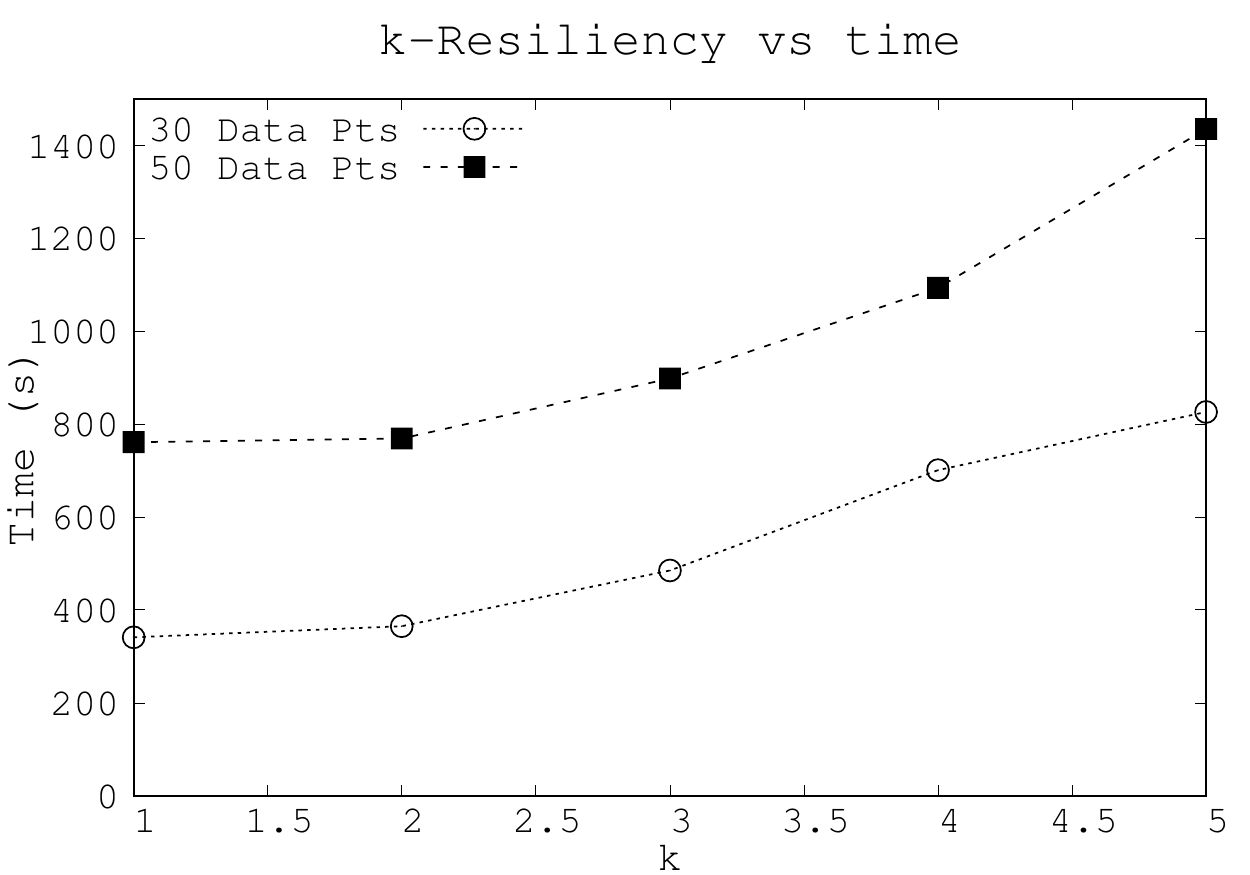}
}
\end{center}
\vspace{-10pt}
\caption{UAVTraj model synthesis time (a) w.r.t. $r-$resiliency, and (b) w.r.t. $k-$resiliency.}
\label{Graph_ScaleUAVTraj2}
\end{figure*}
\subsection{Scalability}
\label{subsec:scale}
We present the scalability of our developed tool based on the formal model discussed earlier in Fig.~\ref{Graph_ScaleUAVTraj} and \ref{Graph_ScaleUAVTraj2}. Fig.~\ref{Graph_PointVsTime} shows the trajectory synthesis time for different sizes of the problem space. We vary the number of points in 3D space and observe the time required to obtain the trajectory plan. We can observe that the required time increases with the increment of the number of points. As we have more points in the problem space, the solver requires to solve more constraints, which leads to the almost linear increment of synthesis time. We can also observe that the UNSAT cases require more synthesis time than the SAT cases.

In Fig.~\ref{Graph_UAVVsTime}, we show the required synthesis time by varying the number of UAVs. We show two cases for two problem sizes: 60 points and 90 points. In these cases, we keep the required percentage of coverage and other parameters constant. We observe that as the number of UAVs increases, the required synthesis time also increases. In the case of 90 points, the number of clauses to satisfy the problem increases quickly with the increment of UAVs. As a result, the time also increases quickly.

We present the trajectory synthesis time by varying the required coverage of the data points in Fig.~\ref{Graph_CovVsTime}. As the requirement to cover more points grows, the synthesis time also increases. For larger problem sets, the time grows quickly, as more constraints are needed to be satisfied.

The required verification time with respect to the number of data points is observed in Fig.~\ref{Graph_DataPtVsTime}. We show the time for two cases: one which is for a total of 50 points, and another is for 80 total points where the UAVs can go. We keep all other parameters constant. For example, the required coverage is kept constant at 80\%, the value for $k$ is 2, while $r$ is kept at 3. We can observe that the verification time increases slowly with the increment of the number of data points for a certain problem size (number of total points). This is because, with the increment of data points, the UAVs need to make sure to visit more points, and the solver needs to solve a more strict set of constraints; however, the number of clauses does not increase drastically.

Next, we present the verification time by varying the value of $r$, which is the minimum number of UAVs that need to visit at least a certain percentage of data points. We can observe that the required time increases quickly as $r$ increases. However, the increment rate is less than the case of increasing $k$. This is because the solver needs to verify less strict conditions when $r$ increases than when $k$ increases.

Fig.~\ref{Graph_kResVsTime} presents the change in verification time with the change of the value of $k$, which is the minimum number of UAVs visiting a certain percentage of the data points while visiting them at a certain time difference.  We keep other parameters, such as the number of UAVs and total points, required coverage, etc., constant and show the results for 30 data points and 50 data points. As the value of $k$ increases, the verification time increases rapidly. With the increment of $k$, the constraints become strict very quickly, as the solver needs to find more complex trajectories for the UAVs so that they can visit more points.

\subsection{Graphical Simulator}
\label{subsec:sim}
We developed a simulator program to verify the effectiveness of the trajectory generation tool. 
The simulator is developed in Unity 3D~\cite{unity}. A user of the simulator can specify the coordinates of the possible waypoints in a three-dimensional space. The simulator reads the number of UAVs and their properties, such as speed, mileage, turning angle constraints, etc., from a user-provided input file. The trajectories are provided based on the output of the trajectory generator. The times of reaching each waypoint on their trajectories, hovering times, etc. are also provided to the simulator. When started, the simulator shows the movement of the UAVs according to its input and proves the effectiveness of the trajectory generator by showing that the UAVs do not collide, but collectively collect the required data and complete the job before the deadline in order to prevent fuel outage, as well as satisfy the resiliency requirements. Appendix~\ref{App_sim} provides further information on this simulator, including a screenshot of the interface.

\section{Conclusion}
\label{Sec:Conclusion}

UAVs are increasingly used in surveillance, as well as data collection from remote sensors. In this paper, we have developed Synth4UAV that can automatically generate the efficient trajectories of a set of UAVs to perform their intended tasks collaboratively. We have considered the routing and maneuvering constraints, budget and time constraints, and the resiliency requirements of the data collection process. We have designed the trajectory planning, which is a combinatorially hard problem, as an SMT-based constraint satisfaction problem, the solution of which generates the trajectories that the UAVs should follow. We have analyzed the relation between different design factors like data coverage, swarm size, budget, resiliency, etc. The evaluation results have proved the scalabiity of Synth4UAV.

\bibliographystyle{unsrt}
\bibliography{aa}
\balance

\newpage
\appendix
\section{Input and Output to the Case Study}
\label{App_input_output}

\begin{table}[h!]
\caption{The input to the case study}
\label{Tab_Example_Input}
\scriptsize
\centering
\begin{tabular}{|p{3.5in}|}
\hline
\vspace{0.025in}

\# Number of waypoints\\
30\\
\# UAV x-coordinates\\
100 500 500 1000 1000 2000 2000 2000 2500 3000 3000 3000 4000 4500 4500 5000 5500 6000 1000 2700 2000 500 2000 3500 5000 4000 500 5500 1500 3500\\

\# UAV y-coordinates\\
100 1000 2000 500 2000 500 1500 2500 2000 1000 1500 2500 1500 1000 2500 2000 1500 2000 3000 3400 4000 3500 2000 3500 3500 3000 3000 500 3500 4000\\

\# UAV z-coordinates\\
0 100 100 200 250 200 150 150 200 200 200 200 200 200 200 200 150 200 200 200 200 200 200 200 200 200 150 100 50 0\\

\# Number of UAVs\\
5\\
\# UAV velocities (m/s)\\
50 48 49 51 52\\
\# UAV mileage (mpg)\\
10 12 11 10 12\\
\# Initial UAV angles (degree with x-axis)\\
0\\
\# Turn angle threshold\\
90\\
\#Climb angle threshold\\
30

\# Source, Destination\\
1 18\\

\# Forbidden points\\
4\\
6 14 19 30\\

\# Points with data sources\\
15\\
2 3 5 9 10 13 20 12 16 18 29 8 26 7 22\\

\# Data coverage threshold (\%)\\
80\\

\# Data freshness threshold (sec)\\
20\\

\# $k-$resiliency\\
3\\

\# Resilient data coverage threshold (\%)\\
70\\

\# Fuel price (\$/gal)\\
3\\

\# Time constraint (sec)\\
2000\\

\# Cost constraint (\$)\\
6000\\
\\ \hline
\end{tabular}
\normalsize
\end{table}

Table~\ref{Tab_Example_Input} lists the input file for the trajectory generator tool for the example case study in Section~\ref{Subsec:case}. The file includes the number of waypoints, the Cartesian coordinates of the points in the three dimensional space, the number of UAVs that will be considered, along with their velocities, mileages, etc. It also includes the initial direction of the UAVs with x axis, the thresholds on turn and climb angles. The source and destination points, the forbidden points which the UAVs should avoid, and the points that are on top or in close proximity of the data sources are also supplied in the input file. The data coverage threshold specifies percentage of the data points that need to covered by the UAVs collectively. The required resiliency specifications are also provided in the input file. The time constraint or the deadline and the available cost constraint or the budget are also supplied.

\begin{table}[h!]
\caption{The output to the case study}
\label{Tab_Example_Output}
\scriptsize
\centering
\begin{tabular}{|p{0.65in}p{0.65in}p{0.65in}p{0.65in}|}
\hline
\vspace{0.015in}
& & &\\ 
\multicolumn{4}{|l|}{\#Required verification time: 336.77}\\
\multicolumn{4}{|l|}{\#We have a solution}\\

\vspace{0.01in}UAV	&\vspace{0.01in}Point	&\vspace{0.01in}Time	&\vspace{0.01in}Hover\\
1	&1	&0	&0\\
1	&2	&20.79898987		&1\\
1	&3	&41.79898987		&0\\
1	&27	&62.82397427		&0\\
\ldots &\ldots&\ldots&\ldots\\
1	&17	&194.898209	&0\\
1	&18	&209.0756559	&0\\
\ldots &\ldots &\ldots &\ldots\\
4	&1	&0	&1\\
4	&4	&19.7056384	&0\\
4	&5	&38.7084393		&1\\
\ldots &\ldots &\ldots &\ldots\\
\vspace{0.015in} & & &\\
\multicolumn{4}{|l|}{\#All trajectories:}\\
UAV	&Src	&Dest	&\\
1	&1	&2	&\\
1	&2	&3	&\\
1	&3	&27	&\\
1	&27	&29	&\\
\ldots &\ldots &\ldots&\\
1	&21	&20	&\\
1	&20	&12	&\\
1	&12	&11	&\\
\ldots &\ldots &\ldots&\\
4	&1	&4	&\\
4	&4	&5	&\\
\ldots &\ldots &\ldots&\\
\vspace{0.015in} & & &\\
\hline
\end{tabular}
\normalsize
\end{table}

Table~\ref{Tab_Example_Output} presents the contents of the output file generated by the trajectory generation tool. As can be seen from the output file, it provides the waypoints on the trajectories of all the UAVs. It also provides the times when the UAVs should reach the waypoints and the time of hovering at those points.

\section{Sample Z3 Code}
\label{App_Code}
Table~\ref{Tab_z3Code} presents a short listing of the z3 code generated by the trajectory generation tool. It shows a few of the expressions that assert the constraints according to our constraints satisfaction model. These expressions are generated for the satisfiable case study shown in Section~\ref{Subsec:case}. For example, the first line in the presented code has two Boolean variables $\texttt{Visit\_1\_18}$ and $\texttt{Visit\_1\_18}$, which denote that UAV $1$ should visit both point $1$ and $18$.
%
\begin{table}[h!]
\caption{Sample Code}
\label{Tab_z3Code}
\scriptsize
\begin{center}
\begin{tabular}{|p{3.25in}|}
\hline
\begin{verbatim}
(assert (and Visit_1_1 Visit_1_18))
(assert (and Visit_2_1 Visit_2_18))
...
(assert (not Visit_2_6))
(assert (not Visit_2_14))
...
(assert (=> Travel_1_1_2 (and Visit_1_1 Links_1_2)))
(assert (=> Travel_1_1_2 
     (= T_1_2 (+ T_1_1 TTmp_1_2 (to_real Hover_1_2)))))
(assert (=> Travel_1_1_2 (= P_1_2 (+ P_1_1 PTmp_1_2))))
...
(assert (=> (and Visit_1_2 Visit_2_2) (or 
     (>= (- T_1_2 T_2_2) 1.0) (>= (- T_2_2 T_1_2) 1.0))))
...
(assert (<= T_1_18 2000.0))
(assert (<= P_1_18 6000.0))
...
\end{verbatim}
\\\hline
\end{tabular}
\end{center}
\vspace{-10pt}
\end{table}

This is asserted by the solver because we have a constraint in equation 10 in Section~\ref{subsec:movement}, which basically states that all the UAVs must visit the starting and ending points. Similarly, line 5 of the code states that if UAV $1$ travels from point $1$ to $2$ ($\texttt{Travel\_1\_1\_2}$), then it must be true that it has already visited point $1$ ($\texttt{Visit\_1\_1}$) and there is a possible path/link from point $1$ to $2$ ($\texttt{Links\_1\_2}$) based on its direction at point $1$ and the distance between points $1$ and $2$.

\begin{figure}
	\centering
	\includegraphics[keepaspectratio,scale=0.55,bb=0 0 762 542]{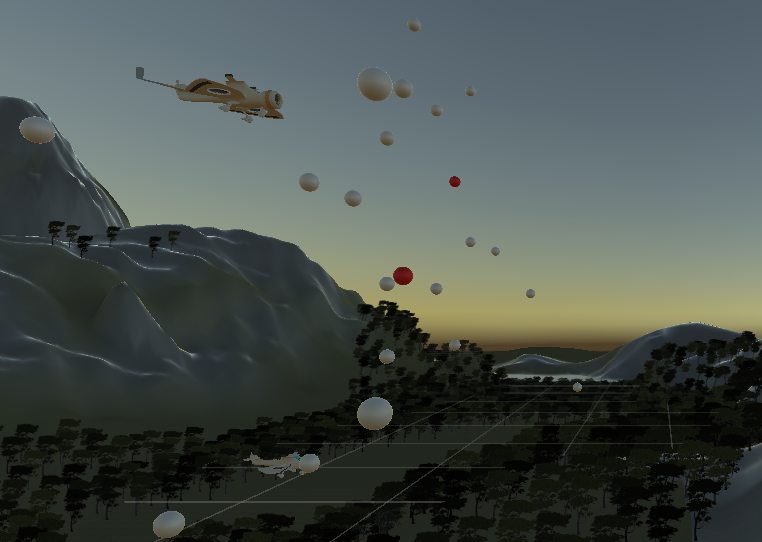}
	\caption{Screenshot of the 3D simulator for UAV trajectories.}
	\vspace{-12pt}
	\label{Fig_Sim}
\end{figure}
\section{Simulator Development}
\label{App_sim}
The simulator has been developed with Unity 3D using C\# programming language\footnote{https://tinyurl.com/y2z8aena}. Fig.~\ref{Fig_Sim} presents a screenshot of the simulator tool that we developed. The figure shows two UAVs flying from point to point collecting data. The spheres represent the waypoints above or in close proximity to the data sources. The red spheres denote the forbidden points which the UAVs avoid. A user can provide the trajectory of one or more UAVs as a combination of points as Cartesian coordinates in a three-dimensional space. The times of reaching each points, as well as the hovering time at the points to avoid collision and proper data collection are also provided from an input text file. The simulator presents an animation of the data collection scenario by the UAVs in a 3D environment. The simulation confirms the effectiveness of the generated trajectories by showing the absence of collisions, as well as the time to reach the end points of the trajectories of the UAVs.

\end{document}